\documentclass[12pt]{amsart}

\usepackage{amsmath}
\usepackage{amsfonts,amssymb,amsthm,enumitem}
\usepackage[margin=1.2in]{geometry}
\usepackage{mathtools}
\usepackage{nicefrac,setspace}
\usepackage{xcolor}
\usepackage{bbold}

\usepackage{tikz}
\usetikzlibrary{patterns}
\usepackage{enumitem}
\usepackage{hyperref}       % hyperlinks
\usepackage{url}            % simple URL typesetting
\usepackage{booktabs}       % professional-quality tables
\usepackage{amsfonts}       % blackboard math symbols
\usepackage{nicefrac}       % compact symbols for 1/2, etc.
\usepackage{microtype}      % microtypography
\usepackage{comment}   %allows for multi-line comments
\usepackage{mathtools}
\usepackage{csquotes}
\MakeOuterQuote{"}
\usepackage{bm}

\makeatletter
\newcommand*\bigcdot{\mathpalette\bigcdot@{.5}}
\newcommand*\bigcdot@[2]{\mathbin{\vcenter{\hbox{\scalebox{#2}{$\m@th#1\odot$}}}}}
\makeatother

\newtheorem{theorem}{Theorem}

\newtheorem*{theorem*}{Theorem}

\newtheorem*{claim*}{Claim}
\newtheorem*{remark*}{Remark}
\newtheorem*{lemma*}{Lemma}
\newtheorem*{comm}{Comments}
\newtheorem{definition}{Definition}
\newtheorem{lemma}[theorem]{Lemma}

\newtheorem*{fact*}{Fact}
\newtheorem{corollary}[theorem]{Corollary}

\newcommand{\N}{\mathbb{N}}

\newcommand{\F}{\mathbb{F}}
\newcommand{\E}{\mathop \mathbb{E}}

\newcommand{\bH}{\mathbf H}
\newcommand{\bA}{\mathbf A}
\newcommand{\bG}{\mathbf G}
\newcommand{\bM}{\mathbf M}
\newcommand{\Cay}{{\sf Cay}}
\renewcommand{\epsilon}{\varepsilon}

\title[Erasure Capacity-Achieving Quantum Tanner Codes]{Achieving Classical Erasure Capacity with Tensor Codes and Quantum Tanner Codes that Achieve Capacity on the Erasure Channel}

\title[Erasure Capacity-Achieving Quantum Tanner Codes]{Achieving Capacity on Erasure Channels with Classical Tensor Codes and Quantum Tanner Codes }

\title[Erasure Capacity-Achieving Quantum Tanner Codes]{Classical Tensor Codes and Quantum Tanner Codes Achieve Capacity on Erasure Channels}

\newif\iffamily
\familyfalse
\newif\ifswap
\swaptrue
\newcommand{\markcolor}{black}

\newif\ifgpt
\gptfalse
\newcommand{\an}[1]{\ifgpt\textcolor{blue}{#1}\fi}

\newcommand{\cF}{\mathcal F}
\newcommand{\be}{\mathbf e}
\newcommand{\bx}{\mathbf x}
\newcommand{\bz}{\mathbf z}
\newcommand{\by}{\mathbf y}
\newcommand{\bc}{\mathbf c}
\newcommand{\br}{\mathbf r}
\newcommand{\bu}{\mathbf u}
\newcommand{\bp}{\boldsymbol{\rho}}

\newcommand{\beps}{\bm \varepsilon}
\newcommand{\supp}{\sf{supp}}
 \newcommand{\bzero}{\mathbf 0}

\begin{document}

\author{Henry D. Pfister$^*$}
\address{$^*$ Duke University, henry.pfister@duke.edu}

\author{Oscar Sprumont$^{\dagger}$}
\address{$^\dagger$ Universit\'e de Bordeaux, oscar.sprumont@u-bordeaux.fr}

\author{Gilles Z{\'{e}}mor$^{\ddagger}$}
\address{$^\ddagger$ Universit\'e de Bordeaux and Institut universitaire de France, zemor@math.u-bordeaux.fr}

\begin{abstract}
We show that on the classical erasure channel, the tensor product of two capacity-achieving code families is itself capacity-achieving under mild conditions.
{\color{\markcolor} We also prove that duals of tensor products of independent uniform random binary linear codes achieve erasure capacity.
Using these local codes, for every erasure probability $\rho\in(0,1/2)$ and every rate gap $\varepsilon>0$, we obtain a family of binary quantum Tanner codes of rate at least $1-2\rho-\varepsilon$.
For fixed $\rho$ and $\varepsilon$, these codes have bounded check weights, minimum distance growing linearly with blocklength, and a linear-time erasure decoder whose decoding error decays exponentially with the blocklength.}

% old
%    We show that on the classical erasure channel, the tensor product of two capacity-achieving codes is itself capacity-achieving. For every $R\in(0,1)$, we then use this result to construct a quantum Tanner code of rate $R$ that achieves capacity on the quantum erasure channel.
    %For every $R\in(0,1)$, we construct a quantum Tanner code of rate $R$ that achieves capacity on the quantum erasure channel. Our construction's decoding time and minimum distance are both linear in the blocklength $n$ of the code, and the probability of a decoding error decays exponentially with $n$. One of our main technical results, which is of independent interest, is that on the binary erasure channel, the tensor product of two capacity-achieving classical codes also achieves capacity.
    %Our construction's decoding time and minimum distance are both linear in the blocklength $N$ of the code, and the probability of a decoding error decays exponentially with $N$.
\end{abstract}

\maketitle

\section{Introduction}

There has been considerable progress during the last few years in the theory of
Quantum Low-Density Parity-Check (QLDPC) codes. Among the most notable results
are the construction of families of asymptotically good QLDPC codes, i.e. whose
dimension and minimum distance both scale linearly in the blocklength $N$
\cite{PK22,LZ22,DinurVidick2023good} and \cite{LH22b,hsieh2025explicit}. These
constructions come with linear-time decoders that are guaranteed to correct a
constant fraction of adversarial errors. An interesting residual question is how
do these decoders behave when confronted with random errors? They can obviously
be guaranteed to correct a constant fraction of random errors, but how large can
this fraction be? In the present paper we are interested in constructing a
variant of quantum Tanner codes \cite{LZ22} that can correct a proportion $p$ of
erasures with a rate $R$ that is arbitrarily close to the quantum erasure capacity
$R=1-2p$.

Quantum Tanner codes are heavily inspired by their classical counterpart. A
classical Tanner code~\cite{T81}, in its bipartite graph version, is a binary
linear code that is constructed
from two ingredients, a regular bipartite graph $(V,E)$ of degree $\Delta$, and two
codes $C_0$ and $C_1$ of length $\Delta$. Bits are indexed by the edge-set $E$ of the
graph and the two inner codes $C_0$ and $C_1$ serve as sets of local
constraints. Codewords of the Tanner code are those functions from $E$ to
$\{0,1\}$ that, restricted to any edge-neighbourhood of a left (respectively right) vertex,
represent a codeword of $C_0$ (respectively $C_1$). When the bipartite graph is
a sufficiently good expander graph, and when the minimum distances of $C_0$ and
$C_1$ are large enough, Tanner codes were famously
shown~\cite{SS96} to be asymptotically good. They also come with a natural
left-right parallel decoder~\cite{zem2001} that can be guaranteed to correct a
constant fraction of errors in linear time. If the Tanner code is chosen to be
too symmetric, namely if we take the inner codes $C_0$ and $C_1$ to be equal,
then we understand that we will not be able to decode close to the capacity of,
say, the erasure channel, because almost every inner code will be saturated with
a number of erasures that exceeds its capacity. This obstacle can be
circumvented however, if we consider a less symmetric version of the Tanner
code, namely by using for $C_1$ a code of rate close to $1$, which puts few
constraints on the overall code, so that the rate of the Tanner code will be
close to the rate $R_0$ of the left inner code $C_0$. A first decoding step
consisting of decoding in parallel every little code $C_0$ will thus bring down
the number of errors to a very small proportion of the block length and the
remaining errors can be dealt with by invoking the worst-case guarantees of the iterative
decoder. 
This strategy was pursued in \cite{bz2002} and such Tanner code families were
shown to be capacity-achieving.
% old
%This strategy was pursued in \cite{bz2002} and such Tanner codes were shown to be capacity-achieving.

Our goal is to pursue a similar strategy in the quantum setting. {\em Quantum
Tanner codes}~\cite{LZ22} are instances of binary CSS codes~\cite{CS96,Ste96b}.
We may think of CSS codes as a pair $(C_X,C_Z)$ of classical codes defined by
two parity-check matrices $\bH_X$ and $\bH_Z$, with the
requirement that $X$-checks must be orthogonal to $Z$-checks. The graph used to
define a classical Tanner code is replaced by a {\em square complex}. This is an
incidence structure $(V,E,Q)$ between a set of vertices, a set of edges, and a set of
squares that are $4$-cycles of the graph $(V,E)$. The square complex is
quadripartite, meaning that vertices are partitioned into four sets,
$V=V_{00}\cup V_{01}\cup V_{10}\cup V_{11}$ and every square is incident to a
vertex from each of the four sets. Qubits are indexed by squares and very much
like in the classical case, the code $C_X$ is defined by local constraints that
require codewords, when restricted to square-neighbourhoods of vertices of
$V_{00}\cup V_{11}$ to coincide with codewords of an inner code $C_{00}$.
Similarly, the code $C_Z$ is defined by local constraints of the
square-neighbourhoods of vertices of $V_{01}\cup V_{10}$. In fact, when
restricting ourselves to two sets of vertices among four, squares become
incident to exactly two vertices and the induced incidence structures
$(V_{00}\cup V_{11},Q)$ and $(V_{01}\cup V_{10},Q)$ are bipartite graphs. The
classical codes $C_X$ and $C_Z$ are nothing but classical Tanner codes defined
on these graphs.

In \cite{LZ22} and subsequent papers \cite{LZsoda2025,LZ2023}, the codes are, for
simplicity's sake,
presented with the most symmetric possible structure. This means that the inner code
constraints of the Tanner code $C_X$ are the same for all vertices, and
similarly for the Tanner code $C_Z$. We shall exhibit a way to generalise the
quantum Tanner code construction so that the two associated classical codes
$C_X$ and $C_Z$ each use two distinct inner codes, one of which has a rate that can
be made arbitrarily close to $1$ like in the classical case. A first round of
decoding will remove a large proportion of the erasures, after which we can rely
on the adversarial error-correcting capabilities of quantum Tanner code to
finish removing the remaining erasures.

% new
For this strategy to work, we need to overcome one serious obstacle. The inner codes of
non-trivial rate and fixed length $\Delta^2$ need small decoding error probabilities at erasure rates close to the capacity limit. However, these inner codes cannot be chosen randomly like in the
classical case, because of the orthogonality requirements of the $X$ and $Z$-
checks. The inner codes are chosen to be duals of tensor products of random
codes of length $\Delta$, and we need to ensure that this family of classical
codes is capacity-achieving. This is not an immediately obvious task, in part
because it is notoriously difficult to evaluate the weight distribution of a
tensor code from that of its component codes.
% old
% For this strategy to work, we need to overcome one serious obstacle. The inner codes of
% non-trivial rate and fixed length $\Delta^2$ need to be close to capacity
% achieving. However, these inner codes cannot be chosen randomly like in the
% classical case, because of the orthogonality requirements of the $X$ and $Z$-
% checks. The inner codes are chosen to be duals of tensor products of random
% codes of length $\Delta$, and we need to ensure that this family of classical
% codes is capacity-achieving. This is not an immediately obvious task, in part
% because it is notoriously difficult to evaluate the weight distribution of a
% tensor code from that of its component codes.

%{\color{\markcolor} To discuss capacity-achieving codes in an intuitive way, we will group codes into families.
%A single code family includes sequences of codes with unbounded length and may also include codes of different rate.
%For example, the informal statement that the family of uniform random linear codes achieves capacity on the erasure channel will be made precise later using Definition~\ref{def:codefamilies}.}
We shall obtain a proof that the family of duals of random tensor codes is capacity-achieving in a
roundabout way. First, we shall prove that tensoring any pair of 
capacity-achieving code families yields a capacity-achieving family for the classical
erasure channel, provided the component error probabilities decay sufficiently fast.
% (Corollary~\ref{cor:tensorfamilies})}.
The way we go about this  consists in decomposing the erasure
channel into applying two successive erasure channels of the required erasure
probabilities. We then modify the probability distribution of one of the erasure
vectors in a way that makes decoding the tensor code straightforward, and compare the expected dimensions of the covered subcodes
under the two erasure distributions. We believe such a method to be of independent interest.

Applied to random component codes, we obtain therefore that the family of random tensor codes
is capacity-achieving for the erasure channel, and we shall prove that the family of their
duals must be capacity-achieving as well. 

\noindent\textbf{Results}

\noindent We state our main results below. For more formal versions of Theorems \ref{thm:tensorIntro} and \ref{thm:qTannerIntro}, see Corollary \ref{cor:tensorfamilies} and Theorem \ref{thm:TannerCapacity} respectively. For a formal definition of code families, see Definition~\ref{def:codefamilies}.

\begin{theorem}\label{thm:tensorIntro}
    {\color{\markcolor}Let $\cF_0$ and $\cF_1$ be two families of codes supporting blocklengths in a common unbounded set of natural numbers.} If the families achieve capacity on the erasure channel and have minimum distance $\omega(\log n)$, where $n$ denotes the blocklength, then the tensor family $\cF_0\otimes \cF_1$ also achieves capacity on the erasure channel.
\end{theorem}
%We note that Theorem \ref{thm:tensorIntro} follows from a more general result about the behavior of arbitrary codes under tensoring (Theorem \ref{tensorcodescapacity}). We also note that o
%One notable consequence of Theorem \ref{thm:tensorIntro} is that the family of random dual tensor codes achieves capacity (see Theorem \ref{thmrandomduals} for the formal statement). %We make use of this result to obtain a family of quantum Tanner codes that achieves capacity on the erasure channel:
\begin{theorem}\label{thm:qTannerIntro}
    Consider any $\rho\in(0,\frac{1}{2})$ and any $\varepsilon>0$. 
    Then there exists a family $\cF$ of quantum Tanner codes of rate $\geq1-2\rho-\varepsilon$  that can decode quantum erasures of rate $\rho$ with failure probability decaying exponentially with the blocklength $N$ and decoding time growing linearly with $N$.
    {\color{\markcolor} For fixed $\rho$ and $\varepsilon$, the degree and the local codes are fixed independently of $N$. The resulting codes have constant stabilizer weights and qubit degrees with minimum distance growing linearly with $N$.}
\end{theorem}

Combining the tensor-product result with product expansion, we prove that duals of tensor products of independent uniform random binary linear codes also achieve the erasure capacity (see Theorem \ref{thmrandomduals} for the formal statement). {\color{\markcolor} This is the key classical ingredient needed for the quantum construction in Sections~\ref{sec:QTC} and~\ref{section:quantumTannercapacity}.}

{\color{\markcolor} Our primary contributions are a capacity-achieving code for the quantum erasure channel within the quantum Tanner framework with linear decoding complexity and classical tensor-product results of independent interest. Concurrent work~\cite{MSTW26} constructs random ensembles of binary QLDPC codes that approach the capacity of the quantum erasure channel, with bounded weight stabilizers and exponentially small decoding error probability.
It does not present a linear-time decoder.}
In Section~\ref{sec:background}, we compare our construction to this and other similar results.

\part{Overview}
\section{CSS codes and the quantum erasure channel}\label{section:CSS}
A binary quantum CSS code $\mathcal{C}$ is defined by two binary matrices named $\bH_X$ and
$\bH_Z$, that both have the same number $N$ of columns and have ranks $r_X$ and $r_Z$,
respectively. Every row of $\bH_X$ should be orthogonal to every row of $\bH_Z$.

The code encodes $k=N-r_X-r_Z$ logical qubits; its Hilbert-space dimension is $2^k$.

The number of logical qubits protected by the quantum code is given by
\[
k=\dim\mathcal{C} = N-r_X-r_Z.
\]
{\color{\markcolor} The matrix $\bH_X$ is the parity-check matrix of a classical code $C_X$, and similarly $C_Z$ is the classical code with parity-check matrix $\bH_Z$.}
The minimum distance $d$ of the quantum code is the quantity $d=\min(d_X,d_Z)$,
{\color{\markcolor} where $d_X$ is the smallest weight of a codeword in $C_Z$ but not in $C_X^\perp$ and
$d_Z$ is the smallest weight of a codeword in $C_X$ but not in $C_Z^\perp$.}

The {\em decoding problem} for errors consists of being given a pair of syndromes
$\bH_X\be_Z^\intercal, \bH_Z\be_X^\intercal$ of two binary vectors
$\be_Z,\be_X\in\F_2^n$ {\color{\markcolor} and returning estimated errors $\be_Z',\be_X'\in\F_2^n$.
The decoder succeeds if the estimated errors satisfy
$\be_Z' \in \be_Z+C_Z^\perp$ and $\be_X' \in \be_X+C_X^\perp$, thus identifying the stabilizer coset of the true error.}

The {\em quantum erasure channel} \cite{grassl1997,bennet1997,dz2013} selects
each of the $n$ coordinate positions with independent probability $p$, which is
the channel parameter. We may define the corresponding erasure vector $\beps\in\F_2^N$.
The channel also chooses two conditionally independent (given the erasure pattern) binary vectors 
$\be_Z,\be_X$ that are uniformly distributed in the space of vectors supported
by the support $\supp(\beps)$ of the erasure vector.
The erasure decoder is then given the erasure vector $\beps$ together with the
syndromes $\bH_X\be_Z^\intercal, \bH_Z\be_X^\intercal$. Decoding succeeds if the
decoder identifies the corresponding error cosets $\be_Z+C_Z^\perp$ and $\be_X+C_X^\perp$.

The capacity of the quantum erasure channel is known to equal $1-2p$ for $0\leq p<1/2$. A family
of codes together with an erasure decoding algorithm is said to achieve the capacity of the erasure channel if it can
produce codes with a rate $R=k/N$ arbitrarily close to $1-2p$ and with a
vanishing decoding error probability.

{\color{\markcolor}
The idea of code families extends naturally to binary CSS codes.
Using Definition~\ref{def:codefamiliescapacity}, we can make this precise by
replacing $\dim C/n$ by the quantum rate $k/N$, replacing $1-\rho$ by
$1-2p$, and using the decoding problem from the erasure pattern and measured
syndrome described above. The decoding error probability averages over the erasures and the Pauli errors supported on the erased
positions.
In this definition, $\varepsilon$ is a rate gap;
the parametrization $p=(1-R)/2-\varepsilon$ used elsewhere corresponds to
a rate gap of $2\varepsilon$.
For the bounded-check-weight and linear-time claims, the degree $\Delta$, the local codes, and the distance are chosen after the target rate and gap.
A suitable choice of the finite local codes can be fixed once and used with a sequence of square complexes.

% old
% The capacity of the quantum erasure channel is known to equal $1-2p$. A family
% of codes together with an erasure decoding algorithm is said to achieve the capacity of the erasure channel if it can
% produce codes with a rate $R=k/N$ arbitrarily close to $1-2p$ and with a
% vanishing decoding error probability. 
}

\section{Comparison with Prior Work}\label{sec:background}

Our construction achieves the erasure capacity with a linear-time decoder within the quantum Tanner framework. More precisely, for every fixed $R\in(0,1)$ and $\varepsilon>0$ such that $p=(1-R)/2-\varepsilon>0$, our codes have constant stabilizer weights and admit decoding from the erasure pattern and measured syndrome in linear time with decoding error probability decaying exponentially in blocklength.
In each of these statements, the implied constants depend only on $R$ and $\epsilon$.

The random ensembles of binary QLDPC codes constructed in~\cite{MSTW26} also approach the capacity of the quantum erasure channel with bounded stabilizer weights and exponentially small decoding error probability. In addition, their construction also approaches the quantum Gilbert--Varshamov bound and achieves the hashing bound for memoryless Pauli channels.  But, the efficient-decoding guarantee in~\cite{MSTW26} concerns a smaller adversarial error radius and the paper does not explicitly claim a linear-time decoder.

% Other closely related results; this is one paragraph.
Earlier work~\cite{LST19} approaches quantum erasure capacity in a restricted range of erasure probabilities using stabilizer generators of polylogarithmic weight. Quantum AEL amplification~\cite{BGG24} gives explicit QLDPC codes with near-Singleton parameters over constant-size folded alphabets. Subsequent work~\cite{BJMST24,JMS26,GJS26} establishes strong list-decoding guarantees for folded QLDPC codes.
In particular, \cite{GJS26} provides near-linear-time syndrome-input list decoding. These folded-symbol guarantees do not by themselves establish unique recovery at capacity for independently erased physical qubits.
% Now, we add citations to practical decoders and simulations results for quantum erasure channels in one big list.
A variety of authors have also focused recently on improving and simulating practical decoders for quantum erasure channels~\cite{CLLD24,GYP24,KO26,YGP25,FLRL26,PGYP26}.

\section{Quantum Tanner codes}\label{sec:QTC}
A quantum Tanner code~\cite{LZ22} is defined by two objects, a {\em square complex} and a
set of inner codes that provide the linear constraints making up the rows of
$\bH_X$ and $\bH_Z$. 

\textbf{The quadripartite Cayley square complex.} It is defined with a group $G$
and two symmetric sets of generators $A\subset G$ and $B\subset G$. Symmetry
here means that $A=A^{-1}$ and $B=B^{-1}$.
For simplicity we take
$\Delta \coloneqq |A|=|B|$. We may actually take $A=B$, but we should differentiate between the
two sets, even if they consist of the same elements of $G$. The sets $A$ and $B$
define two Cayley graphs, the left Cayley graph $\Cay(A,G)$ (whose edges consist
of pairs of the form $\{g,ag\}, g\in G, a\in A$) and the right Cayley graph $\Cay(G,B)$ (whose
edges consist of pairs of the form $\{g,gb\}, g\in G, b\in B$). Those two graphs
should be good expander graphs.

The vertices of the square complex are partitioned into four copies of the group
$G$ named $V_{00},V_{10},V_{11},V_{01}$. So a vertex is a couple $(g, ij)$,
$g\in G$, $i,j\in\{0,1\}$.
Squares (or quadrangles) are defined as quadruples of vertices of the form
\ifswap
{\color{\markcolor}
\[
\{ (g,00), (ag, 10), (agb,11), (gb,01)\}
\]}
\else
\[
\{ (g,00), (ag, 01), (agb,11), (gb,10)\}
\]
\fi
as represented on Figure~\ref{fig:square}.

\ifswap
\begin{figure}[h!]
{\color{\markcolor}
\begin{center}
\begin{tikzpicture}
\draw[thick] (0,0) -- (0,2) -- (2,2) -- (2,0) -- (0,0);
\node[below left] at (0,0) {$g \in V_{00}$};
\node[above left] at (0,2) {$ag \in V_{10}$};
\node[above right] at (2,2) {$agb \in V_{11}$};
\node[below right] at (2,0) {$gb \in V_{01}$};
\end{tikzpicture}
\end{center}
\caption{Square $\{ (g,00), (ag, 10), (agb,11), (gb,01)\}$ of the complex.\label{fig:square}
}}
\end{figure}
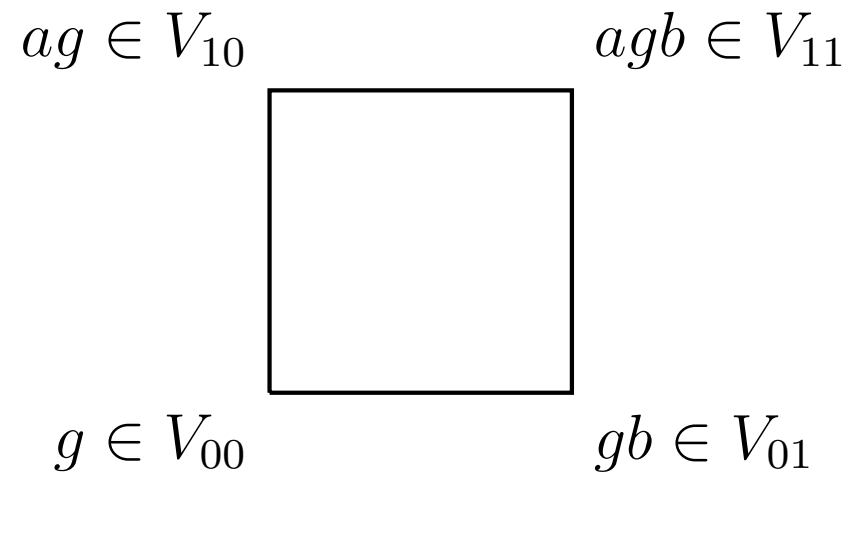
\else
\begin{figure}[h!]
\begin{center}
\begin{tikzpicture}
\draw[thick] (0,0) -- (0,2) -- (2,2) -- (2,0) -- (0,0);
\node[below left] at (0,0) {$g \in V_{00}$};
\node[above left] at (0,2) {$ag \in V_{01}$};
\node[above right] at (2,2) {$agb \in V_{11}$};
\node[below right] at (2,0) {$gb \in V_{10}$};
\end{tikzpicture}
\end{center}
\caption{Square $\{ (g,00), (ag, 01), (agb,11), (gb,10)\}$ of the complex.\label{fig:square}
}
\end{figure}
\fi

\textbf{Local views.} For a vertex $v$, we denote by $Q(v)$ the {\em square
neighbourhood} (or local view) of $v$, i.e. the set of squares incident to $v$. The set $Q(v)$
is naturally isomorphic to $A\times B$ (and hence of cardinality $\Delta^2$).
Given any square $q\in Q$, we may consider the four vertices incident to $q$ and
how their respective local views intersect. Figure~\ref{fig:local} illustrates
the situation. Given a vertex $v$ in $V_{ij}$ and a neighbour $w$ of $v$ in
$V_{(i+1)j}$, we have that the local views $Q(v)$ and $Q(w)$ intersect in
exactly one row of $Q(v)$ and $Q(w)$. If $w$ is a neighbour of $v$ in
$V_{i(j+1)}$, then their local views intersect in exactly one column.
Furthermore, this common row or column is labelled (by elements of $B$ or by
elements of $A$) in exactly the same way, whether it is viewed from $v$ or from
$w$~\cite{LZsoda2025,LZ2023}.

\textbf{The inner codes.}
The $N=|G|\Delta^2$ squares of the square complex index the coordinates (the
qubits) of the quantum code. The local views of the vertices $V_{00}$ and the vertices $V_{11}$
support the $X$-check equations (or $X$-generators) that make up the rows of
$\bH_X$. The local views of the vertices $V_{10}\cup V_{01}$ likewise host
the $Z$-check equations.

In the original quantum Tanner code construction, for simplicity's sake, only two component codes of length
$\Delta$ were used, $C_A$ and $C_B$. Here we shall need a slightly more general
setting involving four component codes of length $\Delta$.

\ifswap
\begin{figure}[h]
{\color{\markcolor}
\begin{center}
\begin{tikzpicture}
\draw (0,0) rectangle (3,3);
\draw[pattern=north west lines,pattern color = blue] (1,0) rectangle (1.5,3);
\draw[pattern=north east lines,pattern color = red] (0,1.5) rectangle (3,2);
\draw[step=0.5cm] (0,0) grid (3,3);

\draw (4,0) rectangle (7,3);
\draw[pattern=north west lines,pattern color = blue] (5,0) rectangle (5.5,3);
\draw[pattern=north east lines,pattern color = red] (5,1.5) rectangle (5.5,2);

\draw[step=0.5cm] (4,0) grid (7,3);

\draw (0,4) rectangle (3,7);
\draw[pattern=north east lines,pattern color = red] (0,5.5) rectangle (3,6);
\draw[pattern=north west lines,pattern color = blue] (1,5.5) rectangle (1.5,6);
\draw[step=0.5cm] (0,4) grid (3,7);

\draw (4,4) rectangle (7,7);
\draw[step=0.5cm] (4,4) grid (7,7);
\draw[pattern=north east lines,pattern color = red] (5,5.5) rectangle (5.5,6);
\draw[pattern=north west lines,pattern color = blue] (5,5.5) rectangle (5.5,6);

\node at (-1.4,0.5) {$Q(g,00)$};
\node at (8.5,0.5) {$Q(gb,01)$};
\node at (-1.4,6.5) {$Q(ag,10)$};
\node at (8.5,6.5) {$Q(agb,11)$};
\node at (1.25,-0.4) {$b$};
\node at (5.35,-0.35) {$b$};
\node at (-0.4,1.75) {$a$};
\node at (-0.4,5.75) {$a$};

\end{tikzpicture}
\end{center}
\caption{The four local views $Q(v)$ that contain the square $\{ (g,00), (ag,
10), (agb,11), (gb,01)\}$. The views of two vertices connected by an $A$-edge
(resp.~a $B$-edge) share a row depicted in red (resp.~a column in blue). We may
choose the labeling so as to ensure that a given square, such as the one in red and blue, is indexed similarly, by $(a,b)$ here, in the four local views. 
\label{fig:local}}
}
\end{figure}
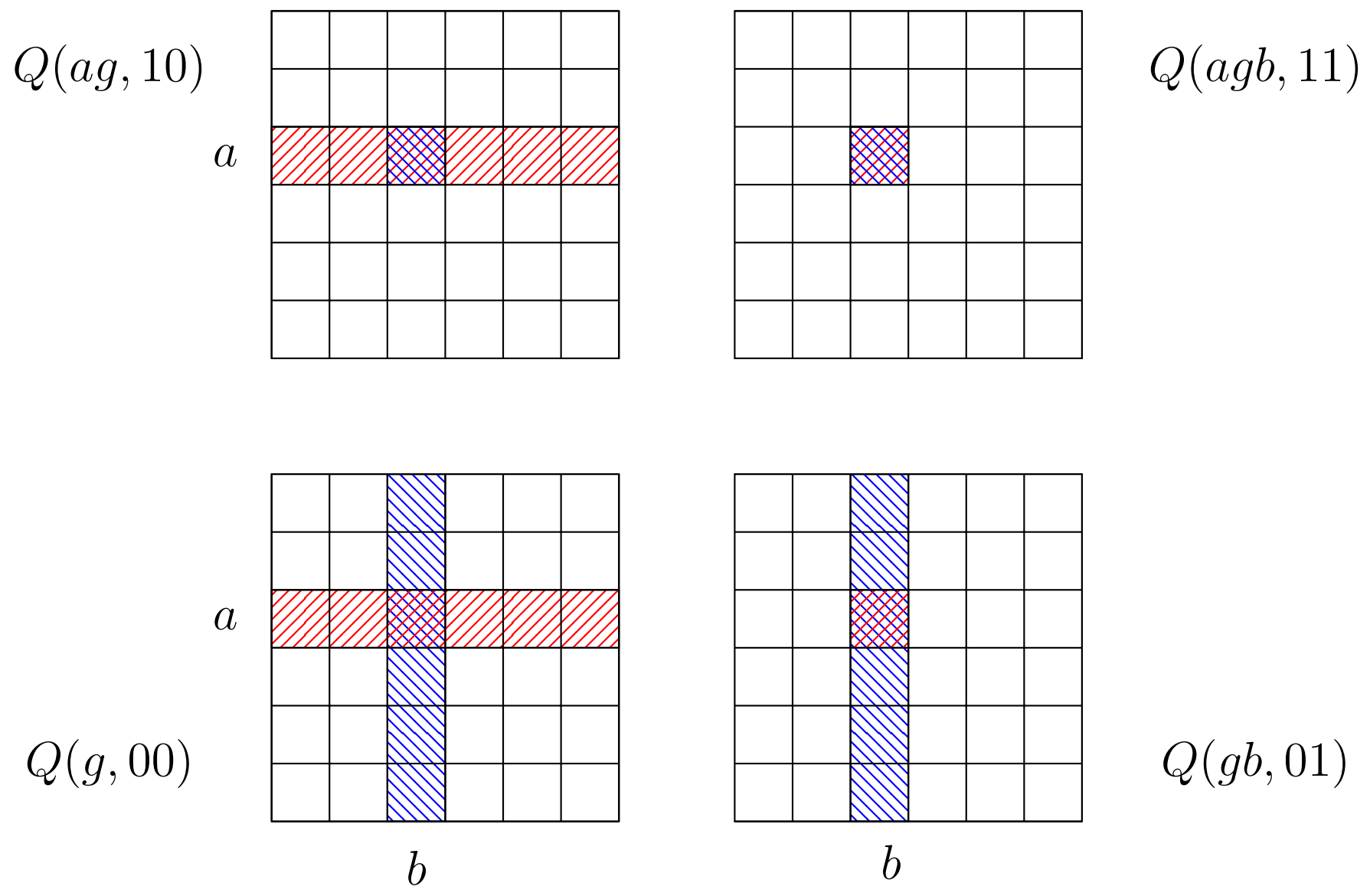
\else
\begin{figure}[h]
\begin{center}
\begin{tikzpicture}
\draw (0,0) rectangle (3,3);
\draw[pattern=north west lines,pattern color = blue] (1,0) rectangle (1.5,3);
\draw[pattern=north east lines,pattern color = red] (0,1.5) rectangle (3,2);
\draw[step=0.5cm] (0,0) grid (3,3);

\draw (4,0) rectangle (7,3);
\draw[pattern=north west lines,pattern color = blue] (5,0) rectangle (5.5,3);
\draw[pattern=north east lines,pattern color = red] (5,1.5) rectangle (5.5,2);

\draw[step=0.5cm] (4,0) grid (7,3);

\draw (0,4) rectangle (3,7);
\draw[pattern=north east lines,pattern color = red] (0,5.5) rectangle (3,6);
\draw[pattern=north west lines,pattern color = blue] (1,5.5) rectangle (1.5,6);
\draw[step=0.5cm] (0,4) grid (3,7);

\draw (4,4) rectangle (7,7);
\draw[step=0.5cm] (4,4) grid (7,7);
\draw[pattern=north east lines,pattern color = red] (5,5.5) rectangle (5.5,6);
\draw[pattern=north west lines,pattern color = blue] (5,5.5) rectangle (5.5,6);

\node at (-1.4,0.5) {$Q(g,00)$};
\node at (8.5,0.5) {$Q(gb,10)$};
\node at (-1.4,6.5) {$Q(ag,01)$};
\node at (8.5,6.5) {$Q(agb,11)$};
\node at (1.25,-0.4) {$b$};
\node at (5.35,-0.35) {$b$};
\node at (-0.4,1.75) {$a$};
\node at (-0.4,5.75) {$a$};

\end{tikzpicture}
\end{center}
\caption{The four local views $Q(v)$ that contain the square $\{ (g,00), (ag,
01), (agb,11), (gb,10)\}$. The views of two vertices connected by an $A$-edge
(resp.~a $B$-edge) share a row depicted in red (resp.~a column in blue). We may
choose the labeling so as to ensure that a given square, such as the one in red and blue, is indexed similarly, by $(a,b)$ here, in the four local views. 
\label{fig:local}}
\end{figure}
\fi

On the local views of vertices of $V_{00}$ we put equations drawn from a basis
of the tensor code $C_A\otimes C_B$, and for vertices of $V_{11}$ we put equations drawn from
a basis of $D_A\otimes D_B$. For vertices of $V_{01}$ we use the code
$C_A^\perp\otimes D_B^\perp$ and for vertices of $V_{10}$ we use the code
$D_A^\perp\otimes C_B^\perp$. The situation is illustrated on
Figure~\ref{fig:localcodes}

\ifswap
\begin{figure}[h!]
{\color{\markcolor}
\begin{center}
\begin{tikzpicture}
\draw[thick] (0,0) -- (0,2) -- (2,2) -- (2,0) -- (0,0);
\node[below left] at (0,0) {$V_{00}: C_A\otimes C_B$};
\node[above left] at (0,2) {$V_{10}:D_A^\perp\otimes C_B^\perp$};
\node[above right] at (2,2) {$D_A\otimes D_B:V_{11}$};
\node[below right] at (2,0) {$C_A^\perp\otimes D_B^\perp:V_{01}$};
\end{tikzpicture}
\end{center}
\caption{Local constraints.}
\label{fig:localcodes} }
\end{figure}
\else
\begin{figure}[h!]
\begin{center}
\begin{tikzpicture}
\draw[thick] (0,0) -- (0,2) -- (2,2) -- (2,0) -- (0,0);
\node[below left] at (0,0) {$V_{00}: C_A\otimes C_B$};
\node[above left] at (0,2) {$V_{01}:C_A^\perp\otimes D_B^\perp$};
\node[above right] at (2,2) {$D_A\otimes D_B:V_{11}$};
\node[below right] at (2,0) {$D_A^\perp\otimes C_B^\perp:V_{10}$};
\end{tikzpicture}
\end{center}
\caption{Local constraints.}
\label{fig:localcodes}
\end{figure}
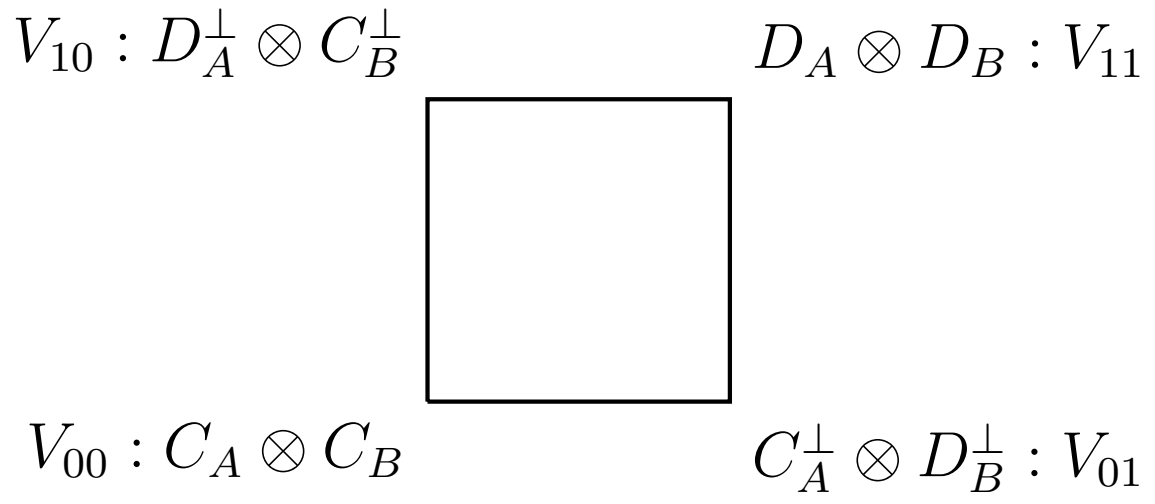
\fi

{\color{\markcolor} We view $A\times B$ as a matrix with rows indexed by $A$ and columns indexed by $B$.
Thus $C_A\otimes C_B$ has columns in $C_A$ and rows in $C_B$. A $B$-edge shares a column and an $A$-edge shares a row. The $V_{00}$--$V_{01}$ and $V_{11}$--$V_{10}$ overlaps pair $C_A$ with $C_A^\perp$ and $D_A$ with $D_A^\perp$, respectively. The $V_{00}$--$V_{10}$ and $V_{11}$--$V_{01}$ overlaps pair $C_B$ with $C_B^\perp$ and $D_B$ with $D_B^\perp$, respectively.
In particular, if $v$ and $w$ are vertices in $V_{00}$ and $V_{01}$ respectively, then their local
views share a column, and an equation on $Q(v)$ hosts a word of $C_A$ on its columns,
while an equation on $Q(w)$ hosts a word of $C_A^\perp$ on its columns, ensuring
orthogonality.
Checking the remaining three case, we see that all $X$-checks therefore commute with all $Z$-checks.}

The original construction used $D_A=C_A$ and $D_B=C_B$, but the extra leeway will be useful to us.

\textbf{Minimum distance and decoding from adversarial errors.} To achieve good distance, we need three properties of the constituents of the quantum Tanner code. First, the square complex should be chosen to have good expansion properties. In more concrete terms, this means that we chose the group $G$ and the generating sets $A$ and $B$ so that associated Cayley graphs $\Cay(A,G)$ and $\Cay(G,B)$ are Ramanujan graphs.
Secondly, we need the component codes $C_A,C_B,D_A,D_B$ and their duals to all have good distance. This means they should be chosen from a family of codes for which the length $\Delta$ can be arbitrarily large, and such that all their minimum distances are at least $\delta\Delta$ for some fixed constant $\delta>0$. This will typically be obtained by chosen the component inner codes to be random of fixed rates.

Finally, the four tensor codes on Figure~\ref{fig:localcodes} should have sufficient {\em robustness} (also called robust testability, or product expansion). 
A pair of codes $C_A,C_B$ of length $\Delta$ has robustness $\kappa$ if for any \textit{dual tensor codeword} $\bx \in (C_A^\perp \otimes C_B^\perp)^\perp$, there exist $\bc \in C_A \otimes \F_2^B, \br\in \F_2^A \otimes C_B$ such that $\bx = \bc+\br$ and
\begin{align*}%\label{eqn:robust}
|\bx| \geq \kappa \Delta (\|\bc\| + \|\br\|),
\end{align*}
where $|\cdot|$ denotes the Hamming weight and $\|\cdot\|$ counts the number of nonzero columns (resp.~rows) of $\bc$ (resp.~$\br$). It was shown \cite{PK22b,KP2025,DinurVidick2023good}, that pairs of random codes of fixed prescribed rates and growing lengths $\Delta$ have a fixed robustness $\kappa$ with a probability that tends to $1$ as $\Delta$ tends to infinity.
When the rates of the inner component codes $C_A,C_B,D_A,D_B$ are fixed, one can satisfy all these requirements simultaneously, with probability tending to one as $\Delta\to\infty$, by choosing independent uniform random linear component codes. 
We have:

\begin{theorem}\cite{LZ2023}\label{thm:LZ2023}
For a constant $\Delta$ large enough, the quantum Tanner codes described in the construction above with a Ramanujan left-right Cayley complex have a linear minimum distance:
\[ d_{\min} \geq \frac{ \delta^2\kappa^2}{256 \Delta} N.\]
\end{theorem}
Theorem~\ref{thm:LZ2023} was proved in \cite{LZ2023} in the special case $D_A=C_A$ and $D_B=C_B$, however the proof extends verbatim to the general case. See Section \ref{section:quantumTannercapacity} for a few further comments.

From \cite[Theorem 13]{LZ2023} we also have:
\begin{theorem}\label{thm:decoding}
For a large enough constant $\Delta$, the quantum Tanner code admits a linear-time decoding algorithm that decodes any error pattern of weight at most $a\delta^2\kappa^3N/\Delta$, for some absolute constant $a$.
\end{theorem}

%It is also proved in \cite{LZ2023} that for fixed $\delta$ and $\kappa$ and large enough $\Delta$, there exists a linear-time decoding algorithm that corrects any error pattern of weight not exceeding a fraction of the length $N$.

\textbf{Asymmetric inner rates.}
Our strategy is to start decoding with two parallel decoding steps, one involving all the $X$-checks on the local views of all the vertices of $V_{00}$, and one involving all the $Z$-checks on the local views of all the vertices of $V_{01}$. We want to claim that these two parallel decoding steps will get rid of a large majority of the erasures, so that the remaining number of erasures falls below the decoding threshold for the worst-case error decoder of \cite{LZ2023}. For this to work, we first need most of the $X$-checks and $Z$-checks to be concentrated on the vertices of $V_{00}$ and $V_{01}$. We proceed in the following way.

{\color{\markcolor} The codes $C_A$ and $C_B$ are chosen with rates $\rho_A$ and
$\rho_B$ respectively. However, the code $D_A$ is chosen with rate approching
$1$, namely $1-\varepsilon$. The code $D_B$ is chosen with vanishing rate
$\varepsilon$.
The total number of $X$ and $Z$ checks are given by
\[
 m_X=N\bigl(\rho_A\rho_B+(1-\varepsilon)\varepsilon\bigr),\qquad
 m_Z=N\bigl((1-\rho_A)(1-\varepsilon)+\varepsilon(1-\rho_B)\bigr).
\]
Consequently $k/N\geq1-(m_X+m_Z)/N$.
Neglecting $\varepsilon$ terms, the total number of $X$ checks is therefore
$\rho_A\rho_BN$ while the total number of $Z$-checks is $(1-\rho_A)N$.

We need the number of $X$-checks to equal the number of $Z$-checks. This is
because we want both sets of parity-check equations to be able to correct the
same erasure patterns and have the same decoding threshold. 
To balance the leading local erasure thresholds as $\varepsilon\to0$, we choose}
\[
1-\rho_A=\rho_A\rho_B=\rho.
\]
So we set $\rho_A=1/(1+\rho_B)$ and $\rho=\rho_B/(1+\rho_B)$. Equivalently,
for every $0\leq\rho\leq1/2$, we set 
\begin{align*}
\rho_A&=1-\rho\\
\rho_B&=\frac{\rho}{1-\rho}
\end{align*}
which give $0\leq\rho_A\leq 1$ and $0\leq\rho_B\leq 1$. For every value
$0\leq\rho\leq1/2$, we achieve therefore a quantum code of rate arbitrarily close to $1-2\rho$. 
%To be precise, not neglecting the $\varepsilon$ terms we obtain a quantum code rate
%\[
%R\geq 1-2\rho-\varepsilon\left(2-\varepsilon-\frac{\rho(2-\rho)}{1-\rho}\right).
%\]
All possible quantum code rates $R$ can therefore be achieved in this way. Note that the rate $\varepsilon$
is chosen to be arbitrarily small, depending on how close we want to be to capacity, but is then fixed and non-zero. A sufficiently large $\Delta$ is then chosen after which the total length can be taken to grow to infinity.

In concrete terms, Theorem~\ref{thm:decoding} translates into:
\begin{corollary}\label{cor:decoding}
Given the values $\rho$ and $\varepsilon$ that determine the above quantum Tanner code construction with a large enough $\Delta$, there is a linear time decoding algorithm that decodes any error pattern of weight at most $f(\rho,\varepsilon)N/\Delta$, where $f(\rho,\varepsilon)$ is a positive quantity that depends only on $\rho$ and $\varepsilon$.
\end{corollary}

The square-neighbourhoods of the vertices of $V_{00}$ make up a partition of the set $Q$ of squares, i.e. the code set of coordinates. For every vertex $v$ of $V_{00}$ we are given a syndrome value $\bH_v\be_v^\intercal$, where $\bH_v$ is the parity-check matrix corresponding to the set of local constraints, which in this case means that $\bH_v$ is a generating matrix for the local tensor code $C_v\simeq C_A\otimes C_B$. We are also given a local erasure pattern $\beps_v$ that supports the error vector $\be_v$. The error vector $\be_v$ is the unique solution corresponding to the syndrome value $s=\bH_v\be_v^\intercal$ whenever the local erasure vector $\beps_v$ does not cover any non-zero codeword of the dual-tensor code $(C_A\otimes C_B)^\perp$ (meaning we do not have $\supp(\bc_v)\subset\supp(\beps_v)$ for any non-zero codeword $\bc_v\in C_v^\perp$). For large enough $\Delta$, unique decoding will occur with overwhelming probability if the erasure probability $p$ is chosen to be below the erasure decoding threshold of the dual tensor code $(C_A\otimes C_B)^\perp$. If the decoding threshold for $(C_A\otimes C_B)^\perp$ can be proved to converge to $\rho_A\rho_B$, for values of $p<\rho_A\rho_B$ and for growing $\Delta$, we recover in this way most of the $Z$ component of the global error vector. Similarly for the $Z$-checks on the square neighbourhoods of vertices of $V_{01}$, for $p<(1-\rho_A)(1-\varepsilon)$ we recover most of the $X$-component of the global error vector. We can then finish decoding by applying the worst-case decoder of 
Corollary~\ref{cor:decoding}.

%{\color{\markcolor} The decoder declares local success only when the erased columns of $\bH_v$ are linearly independent. Otherwise it leaves the whole local block unresolved. Because local views within either selected vertex type partition the qubits, the unresolved weight in each Pauli component is at most $\Delta^2$ times the number of failed local blocks. After fixing the local codes, these block-failure indicators are independent within each selected vertex type. A Chernoff bound then reduces the global failure probability to $\exp(-\Omega(N))$ once the local failure probability is sufficiently small compared to $1/\Delta$.}

We see that for this method to yield quantum erasure capacity achieving code families, we encounter one obstacle. The component codes $C_A,C_B,D_A,D_B$ are chosen randomly so as to achieve the required distance and robustness properties: does the family of duals of tensors of random classical linear codes achieve capacity for the classical erasure channel?
% old
% We see that for this method to yield quantum erasure capacity achieving codes, we encounter one obstacle. The component codes $C_A,C_B,D_A,D_B$ are chosen randomly so as to achieve the required distance and robustness properties: do the duals of tensors of random classical linear codes achieve capacity for the classical erasure channel?

%We see that for large $\Delta$, a first round of decoding with the $X$-checks
%hosted by $V_{00}$ will lift an overwhelming majority of erasures for erasure
%probabilities close to $\rho$ (at capacity) and a first round of decoding with
%the $Z$-checks at $V_{01}$ will decode an overwhelming majority of erasure patterns of
%weight close to $\rho n$.

\section{Tensor codes and their duals on the classical erasure channel} 

For the construction of quantum Tanner codes we described above, we need to prove that the family of dual tensor codes for two random linear component codes % if $C_1$ and $C_2$ are random linear codes of constant rates, the code $(C_1\otimes C_2)^\perp$ %dual of the tensor product of two uniformly random linear codes 
achieves capacity on the erasure channel. 
Since this is a purely classical question, and in order not to confuse any reader that may only want to read the classical portion of this paper, we revert to classical notation and denote our two component codes by $C_1,C_2{\color{\markcolor}\subseteq}\F_2^n.$ For our application to quantum Tanner codes, the blocklength $n$ will be $\Delta$ and the codes $C_1,C_2$ will be various combinations of $C_A,C_B,D_A,D_B$ and their duals.

We will first prove a much more general statement in Theorem~\ref{tensorcodescapacity}, which is itself of independent interest: that on the erasure channel, the tensor product of any two capacity-achieving code families is capacity-achieving, provided their error probabilities
decay fast enough. %We give a very high-level overview of the proof before outlining it in more details. 
Our proof of this fact is outlined in Sections \ref{subseccorr} and \ref{subsectensor}. %\ref{subsecdual}.%this section. %in the next few subsections.
At a high level, we first show that %, in some sense% that we will make precise in a moment,
introducing a certain type of correlations between erasures can only make the decoding harder. %(see subsection \ref{subseccorr}). 
Introducing correlations in accordance with the tensor structure of our code (i.e. correlating entries along either the rows or the columns),
%We then leverage this result to 
we are then able to show that if two arbitrary linear codes $C_1$ and $C_2$ can correct random erasures of rates $\rho_1$ and $\rho_2$ respectively, the code $C_1\otimes C_2$ must be able to correct random erasures of rate $\rho_1+\rho_2-\rho_1\rho_2$, with an error bound
depending on the component errors and lengths. % (see subsection \ref{subsectensor}). 
For capacity-achieving code families with this error decay, this implies that their tensor-product family is capacity achieving. For our application to quantum Tanner codes, we would then want to claim that the family of codes $(C_1\otimes C_2)^\perp$ must achieve capacity as well. %For arbitrary capacity-achieving codes $C_1$ and $C_2$, %we are not able to prove a statement quite as strong; 
%we can only prove that an overwhelming majority of the coordinates can be decoded successfully. 
We show that that is indeed the case as long as $C_1$ and $C_2$ satisfy the robustness property; since uniformly random linear codes are robust with high probability, we get that the family of dual tensor products of two random linear codes achieves capacity. We give a more detailed overview of the proof below.

\subsection{Correlations can only hurt}\label{subseccorr}
We compare the decoding performance of an arbitrary Boolean linear code $C$ under independent and correlated erasures. Fix some parameter $\rho\in(0,1)$ and consider the following
%The main technical result we establish to prove this claim states that, in some sense, introducing correlations between erasures can only make the decoding harder. %Specifically, denoting by $S_C(x)$ the subcode of $C$ that is covered by the erasure pattern $x$, we show that if $x,\tilde{x}$ are two 
%Specifically, denote by $S_C(x)$ the subcode of $C$ that is covered by the erasure pattern $x$ and consider the following 
two randomized erasure patterns: %$x,\tilde{x}$:
\begin{enumerate}
    \item $\bx$ is $\rho$-noisy, meaning that each coordinate is independently sampled to be $1$ with probability $\rho$ and $0$ otherwise.%, for some $\rho\in(0,1)$
   % \item $\tilde{x}$ has $\Pr[\tilde{x}_i=1]=\rho$ for all $i$, but some coordinates are perfectly positively-correlated (that is, some pairs of coordinates $(i,j)$ satisfy $\tilde{x}_i=\tilde{x}_j$).
%    \item $\tilde{x}$ has $\Pr[\tilde{x}_i=1]=\rho$ for all $i$, but some coordinates are sampled simultaneously (that is, some pairs of coordinates $(i,j)$ satisfy $\tilde{x}_i=\tilde{x}_j$).
    \item $\tilde{\bx}$ is $\rho$-noisy, conditioned on some tuples of coordinates being equal (that is, some subsets of coordinates $\{i_1,i_2,\dotsc,i_t\}$ satisfy $\tilde{\bx}_{i_1}=\tilde{\bx}_{i_2}=\dotsc=\tilde{\bx}_{i_t}$).
\end{enumerate}
%Note that for any linear code $C\subseteq\F_2^n$ and any erasure pattern $x\in\F_2^n$, the number of codewords $c\in C$ whose support satisfies $\textnormal{supp}(c)\subseteq\textnormal{supp}(x)$ is exactly equal to the number of codewords of $C$ that cannot be distinguished from the sent codeword. We thus take the dimension of the covered subcode 
We show that the expected dimension of the subset of codewords $c\in C$ covered by the correlated erasure pattern $\tilde{\bx}$ is always greater than the expected dimension of the subset of codewords $c\in C$ covered by the $\rho$-noisy erasure pattern $\bx$. That is, defining $S_C(\bx)$ to be the set of codewords $\bc\in C$ whose support satisfies $\textnormal{supp}(\bc)\subseteq\textnormal{supp}(\bx)$, we have
%$Denote by $S_C(z)$ the subcode of $C$ that is covered by an erasure pattern $z$. We show that
\begin{align}\label{Edim}
    \E_{\bx}[\textnormal{dim }S_C(\bx)]\leq \E_{\tilde{\bx}}[\textnormal{dim }S_C(\tilde{\bx})].
\end{align}
Note that equation \eqref{Edim} is a meaningful proxy for the decoding performance of $C$ under noise $\bx$ and $\tilde{\bx}$, because by linearity of $C$, the number of codewords $\bc\in C$ whose support satisfies $\textnormal{supp}(\bc)\subseteq\textnormal{supp}(\bx)$ is exactly equal to the number of codewords that cannot be distinguished from the sent codeword when corrupted by $\bx$. 

The reason inequality \eqref{Edim} holds is simply that the function $f(\bx):=\dim S_C(\bx)$ is supermodular: $f(\bx)+f(\by)\leq f(\bx\vee\by)+f(\bx\wedge\by)$. With the other coordinates fixed, its expectation under independent Bernoulli($\rho$) coordinates is convex as a function of the common parameter $\rho$. Thus intuitively, a probability distribution that favors strings of extremal weights will have higher expectation than a probability distribution that is concentrated on strings of relative weight $\approx \rho $ (assuming the average relative weight of both random strings is $\rho $).

\subsection{The tensor code $C_1\otimes C_2$}\label{subsectensor}

We will use \eqref{Edim} to prove that if $C_1,C_2\subseteq\F_2^n$ can decode random erasures of rates  $\rho_1$ and $\rho_2$ respectively, then $C_1\otimes C_2$ can decode random erasures of rate $\rho_1+\rho_2-\rho_1\rho_2$, with failure probability at most $n^2(n\delta_1+\delta_2)$,
where $\delta_i$ is the failure probability for $C_i$.
%This follows from %The implication for capacity-achieving families, including the required error decay, is given in Corollary~\ref{cor:tensorfamilies}. %Suppose without loss of generality that the sender sends the $\vec{0}$ codeword.
% old
% We will use \eqref{Edim} to prove that if $C_1,C_2\subseteq\F_2^n$ can decode random erasures of rates  $\rho_1$ and $\rho_2$ respectively, then $C_1\otimes C_2$ can decode random erasures of rate $\rho_1+\rho_2-\rho_1\rho_2$. %Suppose without loss of generality that the sender sends the $\vec{0}$ codeword. 

We think of every codeword of $C_1\otimes C_2$ as an $n\times n$ Boolean matrix whose columns are codewords of $C_1$ and whose rows are codewords of $C_2.$ We think of the ($\rho_1+\rho_2-\rho_1\rho_2$)-noisy string $\bx\in\F_2^{n\times n}$ as 
\begin{align}\label{decomposex}
\bx=\by\vee \bz
\end{align}
for a $\rho_1$-noisy string $\by$ and a $\rho_2$-noisy string $\bz.$ (We leave it to the reader to verify that $\bx$ is indeed ($\rho_1+\rho_2-\rho_1\rho_2$)-noisy.) We also define $\tilde{\bz}$ to be a $\rho_2$-noisy $n\times n$ string, conditioned on the fact that entries that are in the same column must be equal %entries that are in the same column being equal
(in other words, each column of $\tilde{\bz}$ is independently sampled to be the all-ones vector with probability $\rho_2$ and the all-zeroes vector with probability $1-\rho_2$).
%\an{HP: Maybe we should define and  these noise models more formally. Also, $\vec{0}$ and $\vec{1}$ for all-zeros and all-ones is a bit odd.}

Equation \eqref{Edim} bounds the expected covered-subcode dimension for $\by\vee \bz$ by that of $\by\vee\tilde \bz$, for each fixed erasure pattern $\by\in\F_2^{n\times n}$.
Thus, using equation \eqref{decomposex}, one can prove that $C_1\otimes C_2$ can decode ($\rho_1+\rho_2-\rho_1\rho_2$)-noisy erasures, it is sufficient to prove that $C_1\otimes C_2$ can decode the erasure pattern $\by\vee\tilde{\bz}$  with sufficiently small failure probability. We thus give a simple decoding algorithm for decoding erasures of type $\by\vee\tilde{\bz}$: first, independently decode each column using the code $C_1.$ Then, decode the remaining erasures row-by-row by using the code $C_2.$ Note that by assumption, $C_1$ can decode random erasures of rate $\rho_1$.
Writing $\delta_1$ for its failure probability, a union bound shows
that all columns can decode their $\by$-erasures with probability at least $1-n\delta_1$.
On this event, every column of the sent message that is not erased by the erasure pattern $\tilde{\bz}$ will be decoded successfully in our algorithm's first step. For every row, our algorithm's second step then only needs to be able to recover the entries that were erased by $\tilde{\bz}$. But restricted to any single row, $\tilde{\bz}$ is just a $\rho_2$-noisy string. Since $C_2$ can decode random erasures of rate $\rho_2$, the second step of our algorithm has conditional failure probability at most
$\delta_2$ on this event, since $\by$ and $\tilde \bz$ are independent.
Thus the two-stage failure probability is at most $n\delta_1+\delta_2$.
Upper bounding the covered-subcode dimension by $n^2$ gives the independent-erasure failure bound $n^2(n\delta_1+\delta_2)$ and the family corollary uses the sufficient condition that this bound tends to zero. %We believe that this approach of introducing correlations to leverage the structure of the code may have useful applications elsewhere.

\subsection{The dual tensor code $(C_1\otimes C_2)^\perp$}\label{subsecdual}

Once we have established that the family of tensor codes $C_1\otimes C_2$ achieves capacity, we want to prove that its dual family achieves capacity as well. We first use standard arguments to show that for any linear code family of rate $R$ that achieves capacity, the dual family performs "pretty well" under random erasures of rate $R$, in that the expected dimension of its covered subcode is sublinear in the blocklength of the code.

% old
% Once we have established that $C_1\otimes C_2$ achieves capacity, we want to prove that its dual code achieves capacity as well. We first use standard arguments to show that for any linear code of rate $R$ that achieves capacity, the dual code performs "pretty well" under random erasures of rate $R$, in that the expected dimension of its covered subcode is sublinear in the blocklength of the code. 

%More specifically, if $C\subseteq\F_2^m$ can with high probability decode random erasures of rate $1-R-\varepsilon$, then
%\begin{align}\label{prettygooddual}
  %  \E_{x\sim R+\varepsilon}[\textnormal{dim }S_{C^\perp}(x)]\lesssim \varepsilon m,
%\end{align}
%where we use the notation $x\sim R+\varepsilon$ to denote a $(R+\varepsilon)$-noisy string $x.$
%Equation (\ref{prettygooddual}) holds because, since $C$ can decode ($1-R-\varepsilon$)-noisy errors, the restriction of its generator matrix $G$ to $(R+\varepsilon)m$ random columns must have full column-rank with high probability. But $G^\intercal$ is the parity-check matrix of $C^\perp$, so by definition the dimension of the covered subcode of $C^\perp$ is the dimension of the kernel of $G^\intercal$ restricted to the rows erased by the erasure pattern. By the rank-nullity theorem, we thus get that when we erase a fraction $(R+\varepsilon)$ of the coordinates, the covered subcode of $C^\perp$ typically has dimension $\lesssim (R+\varepsilon)n-Rn=\varepsilon n$ with high probability.% with high probability.
%Once we have proven \eqref{prettygooddual} for some $\varepsilon=o(1)$, it remains to show that by 

We then want to show that randomly revealing a sublinear number of erased coordinates (or equivalently, lowering the rate of erasures from $R$ to $R-o(1)$) will decrease the expected dimension of the covered subcode from sublinear to $o(1)$. To show this, it is sufficient to show that, with high probability over the choice of the $R$-noisy string $\bx$, the weight of every codeword of $(C_1\otimes C_2)^\perp$ that is covered by $\bx$ is linear in the blocklength of the code. Indeed, we only need to reveal a small (barely superconstant) number of random coordinates to uncover a codeword of linear length with high probability. So, every time we %then a union bound over a basis of the covered subcode (which as we established has sublinear dimension) shows that 
reveal a random set of coordinates of barely superconstant size, we expect the dimension of the covered subcode to decrease by at least $1.$ Repeating this process until the dimension of the covered subcode reaches $0$, we get that if the covered subcode had sublinear size to begin with, revealing a sublinear number of random coordinates recovers every erased codeword with high probability.

All that remains to show is then that with high probability, every codeword of $(C_1\otimes C_2)^\perp\subseteq\F_2^{n\times n}$ that is covered by the $R$-noisy erasure pattern $\bx$ has weight linear in the blocklength $n^2$. For this, we make use of the robustness property of our code: every codeword of weight $o(n^2)$ can be expressed as a sum of $o(n)$ column-codewords of $C_1^\perp$ and of $o(n)$ row-codewords of $C_2^\perp.$ Thus if a codeword of $(C_1\otimes C_2)^\perp$ of weight $o(n^2)$ is erased, there must be a column where a $C_1^\perp$-codeword is erased in all but $o(n)$ coordinates (if this is not true, then the same must be true for a row and the code $C_2^\perp$). But removing a sublinear number of rows barely changes the rate of the column code, so since $C_1^\perp$ is a uniformly random code of rate $<1-R$, 
the probability that a column codeword of $C_1^\perp$ is erased in all 
but $o(n)$ coordinates under random erasures of rate $R$ is bounded by $2^{-\Omega(n)}.$ %Taking a union bound over the location of the $o(n)$ rows and columns then finishes the argument.

\an{\textbf{ChatGPT Comment.}
A union bound over a basis does not eliminate every nonzero linear combination: a revealed coordinate can be nonzero in two basis vectors and zero in their sum. Also, the sum-code components of $(C_1\otimes C_2)^\perp$ are $C_1^\perp$ columns and $C_2^\perp$ rows. The proof establishes a weight bound $n^{2-\tau}$ for fixed $\tau>0$, rather than a fixed positive fraction of $n^2$. Replace the two paragraphs starting ``We then want'' and ``All that remains'' by:
\begin{quote}
We next reveal a sublinear number of coordinates in independent batches. Product expansion shows that, with high probability, every nonzero codeword covered by a slightly above-capacity erasure pattern has weight at least $n^{2-\tau}$, for a suitable fixed $\tau>0$. Whenever the remaining covered subcode is nonzero, its support is therefore large, and the next revealing batch hits that support with high probability. Hitting a coordinate on which some remaining codeword is nonzero strictly decreases the subcode dimension. A union bound over the successive dimension decreases proves that the covered subcode becomes zero.
To establish the weight bound, write a low-weight word of $(C_1\otimes C_2)^\perp$ as the sum of a small number of columns from $C_1^\perp$ and rows from $C_2^\perp$. Outside the intersections of these rows and columns, there is no cancellation. Thus erasing the word forces an erasure-decoding failure in a punctured constituent dual code. At an erasure probability near $\operatorname{rate}(C_1)\operatorname{rate}(C_2)$, each such constituent has a fixed positive capacity margin. Random-code bounds and a union bound over the possible row and column supports exclude these failures with exponentially high probability in $n$.
\end{quote}
}

\begin{comm}
    For simplicity, we presented our results for classical tensor codes in the setting where the codes $C_1,C_2$ are binary and have the same blocklength $n$. However, our proofs carry over to $q$-ary codes of different blocklengths (see Theorem \ref{tensorcodescapacity}).
\end{comm}

%{\color{red} The comment above may not be needed if we address those in the recap below. Also, I'm not 100\% sure if below is the best place to place the "slightly more formal" version of our results or if it would make more sense to break them up in different parts of the overview.}

 Before moving on to the formal proofs, we recapitulate the main results of this paper. Points (1), (2) and (3) are stated more formally in Corollary \ref{cor:tensorfamilies}, Theorem \ref{thmrandomduals} and Theorem \ref{thm:TannerCapacity} respectively.

\begin{enumerate}
\item Consider any two families $\cF_0$ and $\cF_1$ of linear codes over any finite field $\F_q$ and the tensor product family $\cF_0 \otimes \cF_1$. % with aspect ratio $\gamma=1${\color{red} (Is aspect ratio needed?)}. {\color{\markcolor} We can probably restrict to square tensor codes and remove aspect. I can do that tonight if y'all agree.}
If the component families achieve capacity on the erasure channel with length-$n$ component error bounds $\delta_0(n),\delta_1(n)$ satisfying
$n^3\delta_0(n)+n^2\delta_1(n)\to0$, then the family of tensor products of a code
from $\cF_0$ and a code from $\cF_1$ can tolerate a proportion $p$ of erasures
arbitrarily close to capacity, meaning to $1-R_0 R_1$.
This implies that the family of tensor products of uniform random linear codes achieves capacity.
\item Duals of tensor codes of independent uniform random linear codes of the same length and rate $R_0$ and $R_1$ can
be decoded from a proportion $p$ of erasures that converges with the code length
to $p=R_0R_1$.
Thus, the duals of random tensor codes achieve vanishing error probability arbitrarily close to capacity (see Theorem~\ref{thmrandomduals}).
\item For any target rate $R$, $0<R<1$ and any $0<\varepsilon<(1-R)/2$, there exists a family
of binary quantum Tanner codes with increasing length $N$ and rate at least $R$, constructed over a left-right Cayley
square complex of fixed degree $\Delta$.
These codes have asymptotically good minimum
distance (i.e., bounded from below by $\delta N$ for a constant $\delta$) and,
when submitted to a quantum erasure channel with erasure probability
$p=\frac{1-R}{2}-\varepsilon$, they admit a linear-time decoding algorithm that achieves a decoding error probability that decreases to zero exponentially with $N$.
Here $\Delta$ and all constants are fixed after $R$ and $\varepsilon$ are chosen. The collection over target rates achieves quantum erasure capacity in the family sense above.

%\begin{theorem}
  %  Consider any $\rho\in(0,\frac{1}{2})$ and any $\varepsilon>0$. 
   % Then there exists a family of quantum Tanner codes $\{ C_{n}\}$ of blocklength $n$ and rate $\geq1-2\rho-\varepsilon$  that can decode quantum erasures of rate $\rho$ with probability at least
   %    $ 1- 2^{-an}$
   % and decoding time at most $ bn$, where the positive constants $a,b$ may depend on $\rho$ and $\varepsilon$ but not on $n.$
%\end{theorem}
\end{enumerate}

\part{Proofs}
\section{Notation and Conventions}
Throughout this paper, we %use $\ln X$ to denote the natural logarithm of a real number $X>0$ and use $\log X$ to denote the logarithm of $X$ in base 2. We 
denote by $[n]$ the set of integers $\{1,2,\dotsc,n\}$ and by $\log X$ the logarithm of a positive real number $X$ in base $2.$ We denote by $\F_q^n$ the set of all row-vectors of length $n$ with entries from $\F_q$, and for any $\bx\in\F_q^n$, we define $\supp (\bx)\subseteq[n]$ to be the subset of indices $i$ where $\bx_i\neq 0$.

A $q$-ary linear code is a subspace $C\subseteq \F_q^n$. We define the rate of a linear code $C\subseteq\F_q^n$ to be $\textnormal{rate}(C)=\frac{\textnormal{dim }C}{n}$ and define its generator matrix and its parity-check matrix to be any matrices $\bG\in\F_q^{\textnormal{dim }C\times n} $ and $\bH\in\F_q^{(n-\textnormal{dim }C)\times n}$ satisfying
\begin{align*}
    C=\left\{\bz\bG:\bz\in\F_q^{\textnormal{dim }C}\right\},
\end{align*}
\begin{align*}
        C=\left\{ \bc\in\F_q^n:\bH\bc^\intercal=\bzero \right\}.
\end{align*}

%where here and in the rest of this paper, we represent elements of $\F_q^n$ by row vectors of length $n$. 
For any matrix $\bM$ with entries in $\F_q$, we define $\|\bM\|_\textnormal{col}$ to be the number of nonzero columns of $\bM$ and define $\|\bM\|_\textnormal{row}$ to be the number of nonzero rows of $\bM$. When it is clear from context, we will sometimes drop the subscript and write $\|\bM\|$ to mean either $\|\bM\|_\textnormal{col}$ or $\|\bM\|_\textnormal{row}$.

For any set $\Omega$ and any probability distribution $\mu$ on $\Omega$, we write $x\sim \mu$ to denote a random variable $x\in\Omega$ that is sampled according to $\mu$. For any $\rho\in[0,1]$ and any $n\in\N$, we abuse notation and write $\bx\sim \rho_n$ to denote a $\rho$-noisy random vector $\bx\in\F_2^n$, i.e. a random vector $\bx\in\F_2^n$ whose $i^\textnormal{th}$ coordinate is (independently across coordinates):
\begin{align*}
    \bx_i=\begin{cases}
1 & \text{with probability $\rho$},\\
0 & \text{otherwise}.
\end{cases}
\end{align*}
When $n$ is clear from context, we remove the subscript and write $\bx\sim \rho$ to mean $\bx\sim \rho_n$.
In the classical case, we will think of the coordinates $i\in[n]$ where $\bx_i=1$ as the coordinates that are erased by the channel and will call $\bx$ the \emph{erasure pattern}.  For any erasure pattern $\bx\in\{0,1\}^n$ and any linear code $C\subseteq\F_q^n$, we define the subcode of $C$ covered by $\bx$ to be
\begin{align*}
S_C(\bx):=\Big\{  \bc\in C:\bc_i=0 \textnormal{ for all }i\in[n] \textnormal{ where }\bx_i=0  \Big\}.
\end{align*}
For simplicity, for any classical code $C\subseteq \F_q^n$ and any erasure pattern $\bx\in\F_2^n$, we denote the indicator function of a decoding success by
\begin{align*}
    \mathsf{decode}(C,\bx)=\begin{cases}
    1 & \text{if $S_C(\bx)=\{\bzero\}$,}\\
   0 & \text{otherwise.}
\end{cases}
\end{align*}
In the quantum setting, we use $\beps$ instead of $\bx$ to denote the erasure pattern. Recall that a CSS code of length $N$ is defined by two Boolean matrices $\bH_X,\bH_Z$ whose rows are orthogonal (see Section \ref{section:CSS}). For any CSS code $\mathcal{C}$ of length $N$ and any erasure pattern $\beps\subseteq \{0,1\}^N$, we define the indicator of a decoding success by
%\begin{align*}
 %   \mathsf{decode}(C,\beps)=\begin{cases}
 %   1 & \text{if $\be_X+C_X^\perp$ and $\be_Z+C_Z^\perp$ are determined by $\beps,\bH_Z\be_X^\intercal,\bH_X\be_Z^\intercal$}\\
 %   \textnormal{ }&\text{for all $\be_X,\be_Z\in\F_2^N$ with $\supp(\be_X),\supp(\be_Z)\subseteq\supp(\beps)$,}\\
  % 0 & \text{otherwise.}
%\end{cases}
%\end{align*}

\begin{align*}
\mathsf{decode}(\mathcal C,\beps)=
\begin{cases}
1& \text{if }S_{C_X}(\beps)\subseteq C_Z^\perp
     \ \text{and}\ S_{C_Z}(\beps)\subseteq C_X^\perp,\\
0&\text{otherwise}.
\end{cases}
\end{align*}
Equivalently, the code $\mathcal{C}$ succeeds in decoding the erasure pattern $\beps$ if the erasure pattern $\beps$ and the syndromes $\bH_Z\be_X^\intercal$ and $\bH_X\be_Z^\intercal$ uniquely determine the cosets $\be_X+C_X^\perp$ and $\be_Z+C_Z^\perp$ for every $\be_X,\be_Z\in\F_2^N$ with $\supp(\be_X),\supp(\be_Z)\subseteq\supp(\beps)$.

\an{\textbf{ChatGPT comment.}
Replace the preceding quantum success display by
\[
\mathsf{decode}(\mathcal C,\beps)=
\begin{cases}
1,& S_{C_X}(\beps)\subseteq C_Z^\perp
     \ \text{and}\ S_{C_Z}(\beps)\subseteq C_X^\perp,\\
0,&\text{otherwise}.
\end{cases}
\]
Add:
\begin{quote}
Equivalently, the erasure pattern and syndromes $\bH_Z\be_X^\intercal$ and $\bH_X\be_Z^\intercal$ determine the cosets $\be_X+C_X^\perp$ and $\be_Z+C_Z^\perp$ for every supported Pauli error. This indicator describes information-theoretic correctability. When discussing a specified algorithm, we state its success event separately.
\end{quote}
Under the convention $C_X=\ker\bH_X$ and $C_Z=\ker\bH_Z$, the two covered-subcode conditions concern $Z$- and $X$-errors, respectively. The distinction between correctability and a particular decoder's success should also be respected in Theorem~\ref{thm:TannerCapacity} and \eqref{PrErr}.}
For any vectors $\bx,\by\in\F_2^n$, we define $\bar{\bx}\in\F_2^n,\bx\vee \by\in\F_2^n$, $\bx\wedge \by\in\F_2^n$ and $\bx\setminus \by\in\F_2^n$ to be the vectors with $i$-coordinate
$$\bar{x}_i=1-x_i,$$
\begin{align*}
[\bx\vee \by]_i=\begin{cases}
    0 & \text{if $\bx_i=\by_i=0$,}\\
   1 & \text{otherwise,}
\end{cases}
\end{align*}
\begin{align*}
[\bx\wedge \by]_i=\begin{cases}
    1 & \text{if $\bx_i=\by_i=1$,}\\
   0 & \text{otherwise,}
\end{cases}
\end{align*}
\begin{align*}
[\bx\setminus \by]_i=\begin{cases}
    1 & \text{if $\bx_i=1$ and $\by_i=0$,}\\
   0 & \text{otherwise.}
\end{cases}
\end{align*}

The following definitions are used to make our informal notion of capacity-achieving code families precise.
\begin{definition}\label{def:codefamilies}
A family $\cF$ of linear codes over a fixed finite field $\F_q$ is a collection
of ensembles (i.e., distributions) over codes with sequences of unbounded length.
Each ensemble $\mu \in \cF$ has a fixed length $n$ but the dimension may be stochastic.
A family supports all rates if, for any $R \in (0,1)$, there is a sequence of ensembles with increasing length and rate converging to $R$ in probability.
A family has fixed-rate $R$ if all sequences of ensembles with increasing length induce rate sequences converging to $R$ in probability.
\end{definition}
\begin{definition}\label{def:codefamiliescapacity}
We say that a family $\cF$ achieves capacity on the erasure channel if, for any rate $R$ supported by $\cF$ and rate gap $\epsilon>0$, the following holds.
For all $\delta>0$, there is an $n_0 \in \mathbb{N}$ and an ensemble $\mu \in \cF$ of length $n_0$ with expected rate at least $R-\epsilon$ and
\begin{align}\label{familycapacity}
%\Pr_{C\sim\mu}
%\left[\frac{\dim C}{n}<1-\rho-\varepsilon\right] +
\Pr_{\substack{C\sim\mu\\x\sim 1-R-\epsilon}}
       [\mathsf{decode}(C,x)=0] &<\delta.
\end{align}
%Here $x\sim\rho$ means that each coordinate is erased independently with
%probability $\rho$, and $\mathsf{decode}(C,x)$ is $1$ if $C$ uniquely corrects
%that erasure pattern and $0$ otherwise.
\end{definition}
\begin{definition}\label{def:tensorcodefamilies}
The tensor-product family $\cF_0\otimes\cF_1$ consists of the distributions of $C_0\otimes C_1$ obtained by drawing $C_0\sim\mu_0\in\cF_0$ and $C_1\sim\mu_1\in\cF_1$ independently. For simplicity, we require the component lengths $n_0$ and $n_1$ to be equal.
For every realization, the rate of the tensor product is the product of the component rates.
We say a code construction achieves capacity when the implied family achieves capacity.
\end{definition}

\section{Correlating Erasures Can Only Hurt}

In this section, we prove a technical tool we will need in order to show that the tensor product of two capacity-achieving code families achieves capacity on the erasure channel under the error-decay hypotheses of
Corollary~\ref{cor:tensorfamilies}.
% old
% In this section, we prove the main technical tool we need in order to show that the tensor product of two capacity-achieving codes achieves capacity on the erasure channel.

 Lemma \ref{correlationshurt} essentially states that, for any linear code $C$, it is easier to decode $C$ from random independent erasures of rate $\rho$ than it is to decode $C$ from random erasures of rate $\rho$ that are correlated in a natural way. We believe this result, as well as our techniques for leveraging it, may have applications elsewhere.

\begin{lemma}\label{correlationshurt}
     Consider any integer $n\in\N$, any finite field $\F_q$, any subset of coordinates $T\subseteq [n]$ and any partition $\{A_1,A_2,\dotsc,A_t\}$ of $ T$. %Consider any erasure parameter $p\in[0,1]$ and any partition $\{A_1,A_2,\dotsc,A_t\}$ of the set $[n]$ of coordinates. Define  Then for any Fix some erasure parameter $p\in[0,1]$ and some partition  and l
    Fix any $\rho\in[0,1]$ and let $\mu$ denote the probability distribution on $\{0,1\}^T$ that sets  
    \begin{align*}
        \bx_{A_i}=\begin{cases}
    (1,1,\dotsc,1) & \text{with probability $\rho$}\\
    (0,0\dotsc,0) & \text{with probability $1-\rho$}
\end{cases}
\end{align*}
independently for every $ i=1,\dotsc,t$. Then for any partial erasure pattern $\bx_{[n]\setminus T}\in\{0,1\}^{[n]\setminus T}$ and any linear code $C\subseteq\F_q^n$, we have
    \begin{align*}
        \E_{\bx_T\sim \rho}\big[\textnormal{dim }S_C(\bx)\big]\leq \E_{\bx_T\sim \mu}\big[\textnormal{dim }S_C(\bx)\big].
    \end{align*}
    
    \end{lemma}
\begin{proof}
    We proceed by induction on $t$. %The base case captures essentially everything.
    \newline
    \noindent
    \textbf{Base case $t=1$:} %At a high level, the base case $t=1$ follows from the fact that the function $f(x)=\textnormal{dim }C_x$ is submodular and the expectation of a submodular function is always convex (that is, the function $F(p):=\E_{x\sim p}[f(x)]$ is convex). We now proceed with the formal proof. The submodularity claim is that for any subsets of coordinates $S,\tilde{S}\subseteq[n]$, we have
    We will first show that the function $f(\bx)=\textnormal{dim }S_C(\bx)$ is supermodular, i.e. that for any $\bx,\by\in\F_2^n$, we have
    \begin{align}\label{submodular}
        \textnormal{dim }S_C(\bx)+\textnormal{dim }S_C(\by)\leq \textnormal{dim }S_C(\bx\vee \by)+\textnormal{dim }S_C(\bx\wedge \by).
    \end{align}
    To see this, note that $S_C(\bx\wedge \by)=S_C(\bx)\cap S_C( \by)$ and that $S_C(\bx\vee \by)\supseteq S_C(\bx)+ S_C(\by)$, and thus
    \begin{align*}
        \textnormal{dim }S_C(\bx\vee \by)&\geq \textnormal{dim }\big(S_C(\bx)+ S_C(\by)\big)\\
        &=\textnormal{dim }S_C(\bx)+\textnormal{dim }S_C(\by)-\left(\textnormal{dim }S_C(\bx)\cap S_C( \by)\right)\\
        &=\textnormal{dim }S_C(\bx)+\textnormal{dim }S_C(\by)-\textnormal{dim }S_C(\bx\wedge \by).
    \end{align*}    
We now show that the independent-erasure expectation is convex along the direction in which all coordinates of $T$ have the same parameter $\rho$, while the coordinates outside $T$ remain fixed. Fix any $\bx_{[n]\setminus T}\in\{0,1\}^{[n]\setminus T}$ and define
%Now that we have established (\ref{submodular}), our proof will follow from the fact that the expectation of any submodular function is convex. Formally, we %fix any partial erasure pattern $x_{[n]\setminus A_t}\in\{0,1\}^{[n]\setminus A_t}$ and 
\an{\textbf{ChatGPT comment.}
In the derivative displays, use partial derivatives $\partial/\partial\rho_i$ and expectations over $\bx_T$, with $\bx_{[n]\setminus T}$ fixed. The multilinear extension need not be jointly convex, so that stronger interpretation should be avoided. In the induction paragraph replace both occurrences of $T'$ by $\tilde T$, and replace ``apply induction again'' by ``apply the base case''. Insert a space in ``(\ref{submodular})to'' and a comma in $\{i_1,\dotsc,i_{|A_t|}\}$.}
\begin{align*}
    F(\rho):=\E_{\bx_T\sim \rho}[\textnormal{dim }S_C(\bx)].
\end{align*}
We claim that $F$ is convex. Indeed, thinking of the parameter $\rho$ as a vector of parameters $\bp=(\rho_{i_1},\dotsc,\rho_{i_{|T|}})$ for the coordinates $\{i_1.\dotsc,i_{|T|}\}=T$, we have
\begin{align*}
    \frac{d}{d\rho}F(\rho)&=\sum_{i\in T}
    \frac{d}{d\rho_i}\E_{\bx\sim \rho}\left[\textnormal{dim }S_C(\bx)\right]\\
    &=\sum_{i\in T}
    \E_{\bx\sim \rho}\big[\textnormal{dim }S_C(\bx\vee \be_i)-\textnormal{dim }S_C(\bx\setminus \be_i)\big]
\end{align*}
and
\begin{align*}
   & \frac{d^2}{d\rho^2}F(\rho)=\sum_{i,j\in T}
    \frac{d}{d\rho_j}
    \E_{\bx\sim \rho}\big[\textnormal{dim }S_C(\bx\vee \be_i)-\textnormal{dim }S_C(\bx\setminus \be_i)\big]\\
&=\sum_{\substack{i,j\in T\\i\neq j}}
    \E_{\bx\sim \rho}\Big[\textnormal{dim }S_C(\bx\vee \be_i\vee \be_j)-\textnormal{dim }S_C(\bx\vee \be_i\setminus \be_j)-\textnormal{dim }S_C(\bx\vee \be_j\setminus \be_i)+\textnormal{dim }S_C(\bx\setminus \be_j\setminus \be_i)\Big].
\end{align*}
Applying (\ref{submodular}) to the strings $\bx\vee \be_i\setminus \be_j$ and $\bx\vee \be_j\setminus \be_i$ then gives
\begin{align*}
    \frac{d^2}{d\rho^2}F(\rho)&\geq 0.
\end{align*}
Thus $F$ is convex, and in particular
\begin{align}\label{Fconvex}
    F(\rho)\leq (1-\rho)F(0)+\rho F(1).
\end{align}
But $$F(\rho)=\E_{\bx_T\sim \rho}\big[\textnormal{dim }S_C(\bx)\big]$$ and $$(1-\rho)F(0)+\rho F(1)=\E_{\bx_T\sim \mu}\big[\textnormal{dim }S_C(\bx)\big],$$ so (\ref{Fconvex}) proves our base case.
\newline
\noindent
\textbf{Induction step:} Suppose our desired claim holds for partitions of size $\leq t-1$ %of any coordinates set, %of any set of coordinates $[m]$ 
and consider a $t$-sized partition $\{A_1,\dotsc,A_t\}$ of the coordinates set $T\subseteq[n].$ Defining the set of coordinates $\tilde{T}=T\setminus A_t$, we have %Decomposing any $x\in\F_2^n$ into $x=(y,z)$ for $y=x_{A_1}x_{A_2}\dotsc x_{A_{t-1}}$ and $z=x_{A_t}$, we have
\begin{align}\label{splitvariables}
    \E_{\bx_{{T}}\sim \rho}\big[\textnormal{dim }S_C(\bx)\big]&=\E_{\bx_{A_t}\sim \rho}\left[\E_{\bx_{\tilde{T}}\sim \rho}\big[\textnormal{dim }S_{C}(\bx)\big]\right].%\\
    %&=\E_{\substack{x_{\tilde{T}}\sim p\\x_{A_t}\sim p}}\big[\textnormal{dim }S_{x_{\tilde{T}}x_{A_t}}(C)\big]\\
    %&=\E_{\substack{x_{\tilde{T}}\sim p\\x_{A_t}\sim p}}\big[\textnormal{dim }S_{x}(C)\big].
\end{align}
Now for any  fixed $\bx_{A_t}\in\{0,1\}^{A_t}$, we can apply induction to the partition $\{A_1,\dotsc,A_{t-1}\}$ of $\tilde{T}$ and the partial erasure pattern $\bx_{[n]\setminus \tilde{T}}=(\bx_{[n]\setminus T},\bx_{A_t}).$ This gives us
\begin{align*}
    \E_{\bx_{{T}}\sim \rho}\big[\textnormal{dim }S_C(\bx)\big]&\leq\E_{\bx_{A_t}\sim \rho}\left[\E_{\bx_{\tilde{T}}\sim \mu}\big[\textnormal{dim }S_{C}(\bx)\big]\right]\\
    &=\E_{\bx_{\tilde{T}}\sim \mu}\left[\E_{\bx_{A_t}\sim \rho}\big[\textnormal{dim }S_{C}(\bx)\big]\right].
\end{align*}
For every fixed $\bx_{\tilde{T}}\in\{0,1\}^{\tilde{T}}$ inside this new expectation, we apply induction again, this time with the trivial partition $\{A_t\}$ of $A_t$ and the partial erasure pattern $(\bx_{[n]\setminus T},\bx_{\tilde{T}}).$ This gives us
\begin{align*}
    \E_{\bx_{{T}}\sim \rho}\big[\textnormal{dim }S_C(\bx)\big]&\leq\E_{\bx_{\tilde{T}}\sim \mu}\left[\E_{\bx_{A_t}\sim \mu}\big[\textnormal{dim }S_{C}(\bx)\big]\right],
\end{align*}
as desired.
\end{proof}

\section{Tensoring Preserves Erasure Capacity}

We are now ready to prove Theorem \ref{tensorcodescapacity}, which states that in some sense, tensoring two classical codes preserves their decoding qualities on the erasure channel. For example, Theorem \ref{tensorcodescapacity} shows that if $C_1$ and $C_2$ are close to capacity-achieving (by which we mean that $C_1$ has rate $1-\rho_1$ and can decode random erasures of rate $\rho_1-\varepsilon$ and $C_2$ has rate $1-\rho_2$ and can decode random erasures of rate $\rho_2-\varepsilon$), then the tensor code $C_1\otimes C_2$ is also close to capacity-achieving: it has rate $1-\rho_1-\rho_2+\rho_1\rho_2$ and can decode random erasures of rate $\rho_1+\rho_2-\rho_1\rho_2-2\varepsilon$. %\varepsilon(2+\varepsilon-\rho_1-\rho_2)$.
%The error bound retains polynomial length factors;
Corollary~\ref{cor:tensorfamilies} gives sufficient conditions for tensor products of families to achieve capacity.

% old
% We are now ready to prove Theorem \ref{tensorcodescapacity}, which states that in some sense, tensoring two classical codes preserves their decoding qualities on the erasure channel. For example, Theorem \ref{tensorcodescapacity} shows that if $C_1$ and $C_2$ are close to capacity-achieving (by which we mean that $C_1$ has rate $1-\rho_1$ and can decode random erasures of rate $\rho_1-\varepsilon$ and $C_2$ has rate $1-\rho_2$ and can decode random erasures of rate $\rho_2-\varepsilon$), then the tensor code $C_1\otimes C_2$ is also close to capacity-achieving: it has rate $1-\rho_1-\rho_2+\rho_1\rho_2$ and can decode random erasures of rate $\rho_1+\rho_2-\rho_1\rho_2-2\varepsilon$.%\varepsilon(2+\varepsilon-\rho_1-\rho_2)$.

\begin{theorem}\label{tensorcodescapacity}
        Let $\rho_1,\rho_2\in[0,1]$ and $\delta_1,\delta_2>0$ be arbitrary and consider any $q$-ary linear codes $C_1\subseteq\F_q^{n_1}$ and $C_2\subseteq\F_q^{n_2}$ such that
        \begin{align*}
            \Pr_{\bx\sim \rho_1}\big[\mathsf{decode}(C_1,\bx)=0\big]\leq \delta_1,
        \end{align*}
        \begin{align*}
            \Pr_{\bx\sim \rho_2}\big[\mathsf{decode}(C_2,\bx)=0\big]\leq \delta_2.
        \end{align*}
        %are resilient to $p_1$-noisy erasures and $p_2$-noisy erasures respectively. 
        Then the tensor code $C:=C_1\otimes C_2$ %is resilient to $(p_1+p_2-p_1p_2)$-noisy erasures.
        satisfies
        \begin{align*}
            \Pr_{\bx\sim \rho}\big[\mathsf{decode}({C},\bx)=0\big]&\leq \E_{\bx\sim \rho}[\textnormal{dim }S_{C}(\bx)\big]\leq (\delta_1 n_2+\delta_2)n_1n_2
        \end{align*}
 for $\rho:=\rho_1+\rho_2-\rho_1\rho_2$.
    \end{theorem}
    \begin{proof}
        Note that the first inequality is trivial, so we need only prove the second one. We think of any codeword $c\in C_1\otimes C_2$ as an $n_1\times n_2$  matrix %$c\in\{0,1\}^{n_1\times n_2}$,
        where each column is a codeword of $C_1$ and each row is a codeword of $C_2.$ We will sample our erasure pattern in two steps: first, we sample a $\rho_1$-noisy erasure patterns $\by\in\{0,1\}^{n_1\times n_2}.$ Then, we sample a $\rho_2$-noisy erasure pattern $\bz\in\{0,1\}^{n_1\times n_2}.$ We  %think of $x,x'$ as erasure patterns in $\{0,1\}^{n_1\times n_2}$ and 
        define our final erasure pattern $\bx=\by\vee \bz\in\{0,1\}^{n_1\times n_2}$ to have an erasure in coordinate $(i,j)$ if at least one of $\by_{ij}$ and $\bz_{ij}$ is $1.$ Note that this final erasure pattern is indeed $(\rho_1+\rho_2-\rho_1\rho_2)$-noisy, i.e.
        \begin{align}\label{decomposep}
            \E_{\bx\sim \rho}[\textnormal{dim }S_{C}(\bx)\big]&=\E_{\by\sim \rho_1}\E_{\bz\sim \rho_2}\big[\textnormal{dim }S_C(\by\vee \bz)\big].
        \end{align}
%where %for convenience we defined 
%we recall our definitions $C:=C_A\otimes C_B$ and $p:=p_1+p_2-p_1p_2$. 
Note also that by Lemma \ref{correlationshurt}, for any fixed erasure pattern $\by\in\{0,1\}^{n_1\times n_2}$ we have
        \begin{align}\label{boundbymu}
        \E_{\bz\sim \rho_2}\big[\textnormal{dim }S_C(\by\vee \bz)\big]\leq \E_{\bz\sim \mu}\big[\textnormal{dim }S_C(\by\vee \bz)\big],
    \end{align}
    where $\mu$ is the probability distribution on $\{0,1\}^{n_1\times n_2}$ which, for every $i\in[n_2]$, independently sets the $i^\textnormal{th}$ column $(\bz_{1i},\bz_{2i},\dotsc,\bz_{n_1 i})$ as follows:
    \begin{align*}
       (\bz_{1i},\bz_{2i},\dotsc,\bz_{n_1 i})=\begin{cases}
    (1,1,\dotsc,1) & \text{with probability $\rho_2$,}\\
    (0,0,\dotsc,0) & \text{with probability $1-\rho_2$.}
\end{cases}
\end{align*}

For fixed $\by$, apply Lemma~\ref{correlationshurt} to $T=\{(i,j):y_{ij}=0\}$, partitioned by its nonempty intersections with the columns. Coordinates outside $T$ are already erased and remain fixed. This gives \eqref{boundbymu}.
   Combining equations (\ref{decomposep}) and (\ref{boundbymu}), we get
    \begin{align}\label{doubleexpectation}
        \E_{\bx\sim \rho}[\textnormal{dim }S_{C}(\bx)\big]&\leq\E_{\by\sim \rho_1}\E_{\bz\sim \mu}\big[\textnormal{dim }S_{C}(\by\vee \bz)\big].
    \end{align}
We now claim that when the erasures are sampled as $\by\vee \bz$ with $(\by,\bz)\sim \rho_1\times\mu$, the following decoder for $C_1\otimes C_2$ succeeds with high probability: first, independently decode each column using maximum-likelihood decoding for the code $C_1.$ Then, decode the remaining erasures row-by-row by using maximum-likelihood decoding for the code $C_2.$ Let $E=E(\by)$ denote the event that every column $j\in[n_2]$ can independently decode its $\by$-erasures using maximum-likelihood decoding for $C_1.$ % decoded  gets decoded correctly in the first step of the algorithm. 
Note that by our theorem's condition on $C_1$ and the union bound, we have
%We now claim that when the erasures are sampled as $x=y\vee z$ with $(y,z)\sim(p_1,\mu)$, the following decoder for $C_A\otimes C_B$ succeeds with high probability: first, independently decode each column using maximum-likelihood decoding for the code $C_A.$ Then, decode the remaining erasures row-by-row by using maximum-likelihood decoding for the code $C_B.$ Let $E=E(y)$ denote the event that every column $j\in[n_2]$  with $z_j=0$ (i.e. every column that is not erased by the random string $z$) gets decoded correctly in the first step of the algorithm. Note that by our theorem's condition on $C_A$ and the union bound, we have
\begin{align}\label{allcolumnsfine}
\Pr_{\substack{\by\sim \rho_1}}[E(\by)]&\geq 1-n_2\delta_1.
\end{align}
Now, note that whenever the event $E$ holds, the first decoding step of our algorithm will successfully recover every column $j\in[n_2]$  with $\bz_j=0$ (i.e. every column that is not erased by the random string $\bz$). Thus conditioned on the event $E$, the second step of our decoding algorithm fails if and only if %the set of columns erased by $z$ contains 
$\bz$ covers a codeword of $C_2.$ By our theorem's condition on $C_2$, this occurs with probability at most
\begin{align*}
    \Pr_{\bz\sim \mu}\big[S_{C}(\by\vee \bz)\neq\{0\}\big|E(\by)\big]&\leq \delta_2.
\end{align*}
Combining this with equations (\ref{doubleexpectation}) and (\ref{allcolumnsfine}) and the fact that dim $S_{C}(\by\vee \bz)\leq n_1n_2$, we get
\begin{align*}
         %\Pr_{x\sim p}\big[\mathsf{decode}(C,x)=0\big]&\leq
         \E_{\bx\sim \rho}[\textnormal{dim }S_{C}(\bx)\big]&\leq
         \E_{\by\sim \rho_1}\E_{\bz\sim \mu}\big[\textnormal{dim }S_{C}(\by\vee \bz)\big]\\
&\leq n_1n_2 \Pr_{\substack{\by\sim \rho_1\\\bz\sim \mu}}\big[\mathsf{decode}(C,\by\vee \bz)=0\big]\\
%\Pr_{y\sim p_1}[E(y)]\E_{z\sim \mu}\big[\textnormal{dim }S_{C}(y\vee z)|E(y)\big]\\
&\leq n_1n_2\left(\Pr_{\by\sim \rho_1}[\neg E(\by)]+\Pr_{\by\sim \rho_1}[E(\by)] \Pr_{\bz\sim \mu}\big[\mathsf{decode}(C,\by\vee \bz)=0|E(\by)\big]\right)\\
&\leq n_1n_2\left( \delta_1 n_2+\delta_2  \right).
\end{align*}

    \end{proof}

As a corollary of Theorem \ref{tensorcodescapacity}, we obtain that under mild assumptions, the tensor product of two capacity-achieving families of codes also achieves capacity on the erasure channel. The following result is a more formal version of Theorem \ref{thm:tensorIntro}. (For any capacity-achieving code families $\cF_0,\cF_1$ with minimum distances $\omega(\log n)$, the decoding error probability of each component code can be brought down to $o(\frac{1}{n^3})$ by decreasing the noise by $o(1)$ (see Theorem \ref{sharptransition}), after which applying Corollary \ref{cor:tensorfamilies} will give a capacity result for the tensor family $\cF_0\otimes \cF_1$).

{\color{\markcolor}
\begin{corollary}\label{cor:tensorfamilies}
Let $\cF_1$ and $\cF_2$ be families of linear codes over the same finite field. Suppose that, for every pair of rates $R_1,R_2\in(0,1)$ supported by
$\cF_1,\cF_2$ and every fixed
$0<\varepsilon<\min\{1-R_1,1-R_2\}$, they contain ensembles
$\mu_{1},\mu_{2}$ of common length $n$ for an unbounded sequence of lengths along which the rates converge in probability to $R_1,R_2$, respectively.
If, for $i=1,2$, we have
%\begin{align}\label{familycomponent}
% &\Pr_{C_i\sim\mu_{i,n}}\left[\frac{\dim C_i}{n}<R_i\right]
% \notag\\
% &\qquad+\Pr_{\substack{C_i\sim\mu_{i,n}\\x\sim1-R_i-\varepsilon}}
%       [\mathsf{decode}(C_i,x)=0]\leq\delta_i(n).
%\end{align}
\begin{align}\label{familycomponent}
 \Pr_{\substack{C_i\sim\mu_{i}\\\bx\sim1-R_i-\varepsilon}}
       [\mathsf{decode}(C_i,\bx)=0]\leq\delta_i(n)
\end{align}
%If $n^3\delta_1(n)+n^2\delta_2(n)\to0$ for each fixed choice of the target
%rates and gap, then $\cF_1\otimes\cF_2$ achieves erasure capacity.
and $n^3\delta_1(n)+n^2\delta_2(n)\to0$, then $\cF_1\otimes\cF_2$ achieves capacity on the erasure channel.% for aspect ratio $\gamma=1$.
%$L=n^2$, and $\rho=1-(R_1+\varepsilon)(R_2+\varepsilon)$. Then
%%\begin{align}\label{tensorfamilyerror}
%% &\Pr_{C_1,C_2}\left[\frac{\dim C}{L}<R_1R_2\right]
%% \notag\\
%% &\qquad+\Pr_{\substack{C_1,C_2\\x\sim\rho}}
%%       [\mathsf{decode}(C,x)=0]
%% \leq n^3\delta_1(n)+n^2\delta_2(n).
%%\end{align}
%{\color{red}
%\begin{align}\label{tensorfamilyerror}
% \Pr_{\substack{C_1,C_2\\x\sim\rho}}
%       [\mathsf{decode}(C,x)=0]
% \leq n^3\delta_1(n)+n^2\delta_2(n).
%\end{align}
%}
In particular, if $\delta_i(n)=O(n^{-a})$ for some $a>3$, the failure rate is
$O((n^2)^{-(a-3)/2})$. If $\delta_i(n)=\exp(-\Omega(n))$, it is
$\exp(-\Omega(\sqrt n^2))$.

%These bounds average over the component ensembles.
\end{corollary}
\begin{proof}
To start, we condition on the component codes and apply
Theorem~\ref{tensorcodescapacity} with their actual decoding error
probabilities.
Averaging the bound gives
\begin{align*}
 \Pr_{\substack{C_1,C_2\\\bx\sim\rho}}[\mathsf{decode}(C,\bx)=0]
 &\leq n^3\Pr_{\substack{C_1\\\bx\sim1-R_1-\varepsilon}}
       [\mathsf{decode}(C_1,\bx)=0]\\
 &\quad+n^2\Pr_{\substack{C_2\\\bx\sim1-R_2-\varepsilon}}
       [\mathsf{decode}(C_2,\bx)=0].
\end{align*}
Then, drawing $C_1,C_2$ independently and defining $C=C_1\otimes C_2$, one can combine this bound with \eqref{familycomponent} to show
\begin{align}\label{tensorfamilyerror}
 \Pr_{\substack{C_1,C_2\\\bx\sim\rho}}
       [\mathsf{decode}(C,\bx)=0]
 \leq n^3\delta_1(n)+n^2\delta_2(n).
\end{align}
\end{proof}
}

\section{Duals of Random Tensor Codes Achieve Capacity}

In this section, we prove the classical result needed for our application to quantum Tanner codes: that the family of dual tensor codes of two random linear codes is capacity-achieving on the erasure channel. We will need a few well-known results, which we state below. The first one is that the family of random linear codes achieves capacity on the erasure channel, a standard fact that, to the best of our knowledge, was first proven in \cite{elias1955erasurecapacity}.
Corollary~\ref{cor:randomtensorfamilies} states the capacity consequences
for the random linear and tensor-product families.
% old
% In this section, we prove the classical result needed for our application to quantum Tanner codes: that the dual tensor code of two random linear codes is capacity-achieving on the erasure channel. We will need a few well-known results, which we state below. The first one is that random linear codes achieve capacity on the erasure channel.%, a standard fact that, to the best of our knowledge, was first proven in \cite{elias1955erasurecapacity}.

\begin{lemma}[\cite{elias1955erasurecapacity}]\label{randomcodescapacity}
    There exists a universal constant $a>0$ such that the following holds. Let $\rho\in(0,1)$ be arbitrary and let $C\subseteq\F_2^n$ be a uniformly random linear code of rate $1-\rho.$ Then for any $\varepsilon\in\left[0,\rho\right]$, we have
    \begin{align*}
        \Pr_{\substack{C\subseteq\F_2^n\\\bx\sim \rho-\varepsilon}}\big[\mathsf{decode}(C,\bx)=0  \big]&\leq 2e^{-a\varepsilon^2n}.
    \end{align*}
\end{lemma}

{\color{\markcolor}
\begin{corollary}\label{cor:randomtensorfamilies}
Let $\cF$ contain the uniform ensemble of binary linear codes of every
length and dimension. Both $\cF$ and $\cF\otimes\cF$ achieve capacity on
the BEC. For equal component lengths $n$, each fixed pair of target rates,
and a fixed positive component gap, the corresponding tensor-product ensemble
has decoding error $\exp(-\Omega(n))$.
\end{corollary}
\begin{proof}
For a target rate $R_i$ and common component length $n$, choose $C_i$
uniformly among subspaces of dimension $\lceil R_i n\rceil$.
Its rate is at least $R_i$ and at most $R_i+1/n$. At erasure probability
$1-R_i-\varepsilon$, the gap to capacity is at least
$\varepsilon/2$ once $n\geq2/\varepsilon$. Lemma~\ref{randomcodescapacity}
therefore bounds its ensemble-average failure probability by
$2e^{-a\varepsilon^2n/4}$.

This establishes capacity for $\cF$ and verifies \eqref{familycomponent} with
exponentially small $\delta_i(n)$. Corollary~\ref{cor:tensorfamilies} gives
capacity for $\cF\otimes\cF$ and the error bound
$2(n^3+n^2)e^{-a\varepsilon^2n/4}=\exp(-\Omega(n))$.
\end{proof}
}

We will also need the fact that in the random-noise model, the decoding error probability of any linear code of superconstant distance has a sharp transition.
\begin{theorem}\label{sharptransition}[\cite{tillich2000sharptransition}]
    There exists a universal constant $a>0$ such that the following holds. For every $n\in\N$ and every linear code $C\subseteq\F_2^n$, if $p^*\in(0,1)$ satisfies $\Pr_{x\sim p^*}[\mathsf{decode}(C,x)=0]=\frac{1}{2}$, then for every $p<p^*$ we have
    \begin{align*}
        \Pr_{x\sim p}[\mathsf{decode}(C,x)=0]\leq e^{-a(p^*-p)^2d_\textnormal{min}(C)}.
    \end{align*}
\end{theorem}

Since this cannot be applied to a ensemble of codes with different thresholds, we apply this theorem only after fixing a code whose failure probability at some $p_+>p$ is at most $1/2$. Monotonicity then gives $p^*\geq p_+$ and hence $\Pr_{\bx\sim p}[\mathsf{decode}(C,\bx)=0]\leq\exp(-a(p_+-p)^2d_{\min}(C))$.

We will also need the fact that random dual tensor codes are robust with high probability. Recall from Section \ref{sec:QTC} that a code $(C_1\otimes C_2)^\perp\subseteq\F_2^{n\times n}$ is said to be \emph{$\kappa$-robust} if any codeword $\bc\in (C_1\otimes C_2)^\perp$ can be expressed as $\bc=\bc_\textnormal{col}+\bc_\textnormal{row}$ for some matrix $\bc_\textnormal{col}\in\F_2^{n\times n}$ whose columns are codewords of $C_1^\perp$, some matrix $\bc_\textnormal{row}\in\F_2^{n\times n}$ whose rows are codewords of $C_2^\perp$, and 
\begin{align*}
\textnormal{wt}(\bc) \geq \kappa n (\|\bc_\textnormal{col}\| + \|\bc_\textnormal{row}\|),
\end{align*}
where $\|\cdot\|$ counts the number of nonzero columns (respectively,~rows) of $\bc_\textnormal{col}$ (respectively,~$\bc_\textnormal{row}$). 
\cite{PK22b} showed that the dual of the product of two random linear codes is $\kappa$-product expanding with high probability, for some constant $\kappa$ depending on the rates of the two component codes.
\begin{theorem}\label{hyperexpansionrandom}\cite{PK22b}
    Let $\rho_1,\rho_2\in(0,1)$ be arbitrary and consider any $n>\frac{8}{\min\{1-\rho_1,1-\rho_2\}}$. Let $C_1,C_2\subseteq\F_2^n$  be two random linear codes of dimensions $(1-\rho_1) n$ and $(1-\rho_2) n$ respectively. Define  
    $$\kappa:=\frac{1}{2}\min\left\{\frac{1}{4}h^{-1}\left(\frac{1-\rho_1}{8}\right)h^{-1}\left(\frac{1-\rho_2}{8}\right),h^{-1}\left(\frac{(1-\rho_1)(1-\rho_2)}{8}\right)  \right\}.$$
   % for $\varepsilon_A:=h^{-1}\left(\frac{\rho_A}{8}\right)$ and  $\varepsilon_B:=h^{-1}\left(\frac{\rho_B}  {8}\right)$. 
     Then we have
     $$\Pr_{C_1,C_2}\left[ (C_1\otimes C_2)^\perp\textnormal{ is $\kappa$-expanding} \right]\geq 1-\varepsilon%5\cdot 2^{-\frac{\rho_A\rho_B}{8}\Delta^2}-16\cdot 2^{-\frac{\rho_A\rho_B}{8}\Delta} .
     $$
     for $\varepsilon=5\cdot 2^{-\frac{(1-\rho_1)(1-\rho_2)}{8}n^2}+16\cdot 2^{-\frac{(1-\rho_1)(1-\rho_2)}{8}n}$.

\an{\textbf{ChatGPT Comment.}
The above formula uses undefined $\rho_A,\rho_B$. More substantially, product expansion of $(C_1\otimes C_2)^\perp$ is expansion of the pair $(C_1^\perp,C_2^\perp)$, whose codimensions are $(1-\rho_i)n$, not $\rho_i n$. The cited proof parametrizes its expansion constant by the codimensions of the pair to which it is applied. Since none of the explicit constants is needed later, replace the theorem body by:
\begin{quote}
Fix $\rho_1,\rho_2\in(0,1)$. There exist $\kappa>0$, $b_0>0$, and $n_0$, depending only on $\rho_1,\rho_2$, such that the following holds for all $n\geq n_0$. If $C_1,C_2\subseteq\F_2^n$ are independent uniform subspaces of dimensions $\lfloor(1-\rho_1)n\rfloor$ and $\lfloor(1-\rho_2)n\rfloor$, then
\[
\Pr_{C_1,C_2}[(C_1\otimes C_2)^\perp\text{ is }\kappa\text{-product expanding}]
\geq1-e^{-b_0n}.
\]
\end{quote}
Use the same constants in the small-codeword proof instead of its repeated undefined formula. Use one term, ``product expanding'', consistently.}
\end{theorem}

Finally, for any linear code $C\subseteq\F_2^n$ and any erasure string $\bx\in\F_2^n$, we will also need a few equivalent expressions for the dimension of the subcode of $ C$ covered by $\bx$, in terms of the generator matrix and the parity-check matrix of $C$. We provide a short proof of the claim.

\begin{lemma}\label{equivalentformulationscoveredsubcode}
    Let $C\subseteq \F_2^n$ be any linear code, let $\bG\in\F_2^{ \textnormal{dim }C\times n}$ be a generator matrix of $C$ and let $\bH\in\F_2^{(n-\textnormal{dim }C)\times n}$ be a parity-check matrix of $C$. Then for any erasure pattern $\bx\in\F_2^n$, we have  
    \begin{align*}
        \textnormal{dim }S_C(\bx)&=\textnormal{dim }C-\textnormal{rank }\bG_{\bar{\bx}}
        =\textnormal{wt}(\bx)-\textnormal{rank }\bH_{\bx},
    \end{align*}
    where for any vector $\by$ and any matrix $\bA$, we denote by $A_{\by}$ the restriction of $\bA$ to the columns indexed by the support of $\by$.
   % \begin{align*}
    %    \textnormal{dim }S_C(x)=\textnormal{wt}(x)-\textnormal{dim }H_{\textnormal{cols}=\bar{x}},
   % \end{align*}
\end{lemma}
\begin{proof}
    Define $R=\frac{\textnormal{dim }C}{n}$. To prove our lemma's first equality, we note that
\begin{align*}
        |S_C(\bx)|%&:=\Big\{c\in C:\textnormal{supp}(c)\subseteq \textnorm{supp}(x)\Big\}\\
        &=\left|\Big\{\bz\in \F_2^{Rn}:[\bz \bG]_{\bar{\bx}}=\bzero\Big\}\right|,
    \end{align*}
    and thus
\begin{align*}
        \textnormal{dim}(S_C(\bx))&=\textnormal{dim ker}( \bG_{\bar{\bx}}^\intercal)\\
        &=Rn-\textnormal{rank }\bG_{\bar{\bx}},
    \end{align*}
    where the second line follows from the rank-nullity theorem. To prove our lemma's second equality, we note that
    
    \begin{align*}
        S_C(\bx)%&:=\Big\{ c\in\textnormal{ker}(H):c_{\bar{x}}=\vec{0} \Big\}\\
        &=\Big\{ \bc\in\textnormal{ker}(\bH):\textnormal{supp}(\bc)\subseteq \textnormal{supp}(\bx) \Big\},
    \end{align*}
    and thus
    \begin{align*}
        \textnormal{dim } S_C(\bx)&=\textnormal{dim ker}(\bH_{\bx})\\
        &=\textnormal{wt}(\bx)-\textnormal{rank }\bH_{\bx},
    \end{align*}
    where again the second line follows from the rank-nullity theorem.
\end{proof}

We now have all the tools we need to prove that the family of dual tensor codes of two random linear codes $C_1,C_2\subseteq\F_2^n$ achieves capacity. Our first step will be to show that at capacity, it is very unlikely for the covered subcode of $(C_1\otimes C_2)^\perp$ to contain a codeword of weight significantly below $n^2.$
% old
% We now have all the tools we need to prove that the dual tensor code of two random linear codes $C_1,C_2\subseteq\F_2^N$ achieves capacity. Our first step will be to show that at capacity, it is very unlikely for the covered subcode of $(C_1\otimes C_2)^\perp$ to contain a codeword of weight significantly below $N^2.$

\begin{lemma}\label{nosmallcodeword}
Fix any $\tau\in(0,2)$ and $\rho_1,\rho_2\in(0,1)$. Then there exists some $n_0=n_0(\tau,\rho_1,\rho_2)$ such that for all $n>n_0$, the dual tensor code $C:=(C_1\otimes C_2)^\perp$ of two random linear codes $C_1,C_2\subseteq\F_2^n$ of rates $(1-\rho_1)$ and $(1-\rho_2)$ satisfies, for $\rho:=\rho_1+\rho_2-\rho_1\rho_2$,
\begin{align*}
\Pr_{\substack{{C_1,C_2}\\x\sim 1-\rho}}\left[ \exists c\in S_C(x):0<\textnormal{wt}(c)\leq n^{2-\tau} \right]%&\leq \frac{1}{a}e^{-an}\\
%&\leq  ae^{-\frac{n}{a}},
&\leq  e^{-an},%5\cdot 2^{-\frac{\rho_A\rho_B}{8}\Delta^2}+16\cdot 2^{-\frac{\rho_A\rho_B}{8}\Delta}+\frac{4\Delta}{\kappa \ln\Delta}e^{-\frac{\varepsilon\Delta}{24\kappa}}\\
%&\leq \frac{6\Delta}{\kappa}2^{-\frac{\rho_A\rho_B}{}}
\end{align*}
where $a=a(\tau,\rho_1,\rho_2)$ is a positive constant  that depends on $\tau,\rho_1$ and $\rho_2$ but not on $n$.

\an{\textbf{ChatGPT comment.}
The dual has erasure-capacity threshold $p_0=(1-\rho_1)(1-\rho_2)$, whereas the statement samples erasures at $1-p_0$. The proof itself switches to $p_0$. Moreover, excluding only the exact weight $n^{2-\tau}$ does not exclude smaller weights (and that exact weight need not even be an integer). Finally, the next theorem uses this lemma at $p_0+\gamma$, not $p_0$. A sufficient corrected replacement for this lemma's body is:
\begin{quote}
Fix $\rho_1,\rho_2\in(0,1)$ and $\tau\in(0,1)$. Set
\[
 p_0=(1-\rho_1)(1-\rho_2),\quad
 g=\min\{(1-\rho_1)\rho_2,(1-\rho_2)\rho_1\},\quad \eta=g/4.
\]
There exist $a>0$ and $n_0$, depending only on $\rho_1,\rho_2,\tau$, such that for all $n\geq n_0$, independent uniform component codes of dimensions $\lfloor(1-\rho_i)n\rfloor$ satisfy, for $C=(C_1\otimes C_2)^\perp$,
\[
 \Pr_{C_1,C_2,\,\bx\sim(p_0+\eta)_{n^2}}
 \left[\exists\,\bc\in S_C(\bx)\setminus\{0\}:\ |\bc|\leq n^{2-\tau}\right]
 \leq e^{-an}.
\]
\end{quote}
The smaller range $0<\tau<1$ suffices for the later choice $\tau=\nu/4$. Monotonicity then covers $p_0+\gamma$ once $\gamma\leq\eta$. R34 gives a proof replacement for this strengthened statement.}
\end{lemma}
% old
% 

\begin{proof}
%Define $\kappa:=\frac{1}{2}\min\left\{\frac{1}{4}h^{-1}\left(\frac{\rho_A}{8}\right)h^{-1}\left(\frac{\rho_B}{8}\right),h^{-1}\left(\frac{\rho_A\rho_B}{8}\right)  \right\}$ and 
Let $\kappa:=\frac{1}{2}\min\left\{\frac{1}{4}h^{-1}\left(\frac{1-\rho_1}{8}\right)h^{-1}\left(\frac{1-\rho_2}{8}\right),h^{-1}\left(\frac{(1-\rho_1)(1-\rho_2)}{8}\right)  \right\}$ and recall from Theorem \ref{hyperexpansionrandom} that
\begin{align}\label{Cexpands}
     \Pr_{C_1,C_2}\left[C\textnormal{ is $\kappa$-expanding} \right]\geq 1-5\cdot 2^{-\frac{(1-\rho_1)(1-\rho_2)}{8}n^2}-16\cdot 2^{-\frac{(1-\rho_1)(1-\rho_2)}{8}n}.
\end{align}
%for $\tilde{\varepsilon}=5\cdot 2^{-\frac{\rho_A\rho_B}{8}\Delta^2}+16\cdot 2^{-\frac{\rho_A\rho_B}{8}\Delta} $.
% We define $\alpha=\frac{\log w}{\log\Delta}-1$ (i.e., we express $w$ in the more convenient form $w=\Delta^{1+\alpha}$) and
But if $C$ is $\kappa$-expanding, every codeword $\bc\in C$ can be decomposed into $\bc=\bc_\textnormal{col} + \bc_\textnormal{row}$ for some $\bc_\textnormal{col} \in C_1^\perp \otimes \F_2^n$ and $\bc_\textnormal{row} \in \F_2^n \otimes C_2^\perp$ 
satisfying
 $$\max\Big\{\|\bc_\textnormal{col}\| ,\|\bc_\textnormal{row}\| \Big\}\leq  \frac{\textnormal{wt}(\bc)}{\kappa n}.$$
  Thus, fix any $w\leq n^{2-\tau}$ and consider any row-set $S\subseteq[n]$ and any column-set $T\subseteq[n]$ of sizes $|S|=|T|=\frac{w}{\kappa n}$. We will show that it is very unlikely for a $(1-\rho)$-noisy erasure pattern $\bx$ to cover any $w$-weight codeword $\bc\in C$ whose rows of $\bc_\textnormal{row}$ are supported on $S$ and whose columns of $\bc_\textnormal{col}$ are supported on $T$. %To this end, we note that in order for  an erasure pattern $x\in\F_2^{\Delta\times \Delta}$ to cover some codeword $c%=c_\textnormal{col}+c_\textnormal{row}
% \in C$ 
%with  
%\begin{align*}
%\textnormal{supp}(c_\textnormal{row})\subseteq S\times[\Delta] 
%\end{align*}
%and
%\begin{align*}
%\textnormal{supp}(c_{\textnormal{col}})\subseteq  [\Delta]\times T. 
%\end{align*}
To this end, we first note that in order for $\bx$ to cover any such codeword $\bc$, we must have $\bx_{ij}=1$ for every $(i,j)\notin S\times T$ where either $[\bc_\textnormal{col}]_{ij}=1$ or $[\bc_\textnormal{row}]_{ij}=1$. But since for every $n$ large enough,
    \begin{align*}
        |S\times T|&=\frac{w^2}{\kappa^2n^2}\leq \frac{w}{\kappa^2n^{\tau}}<\textnormal{wt}(\bc),
    \end{align*}
the support of $\bc$ cannot be contained in $S\times T$. 
    Denoting by $\tilde{C_1}$ and $\tilde{C_2}$ the restrictions of $C_1^\perp,C_2^\perp$ to the coordinate sets $[n]\setminus S$ and $[n]\setminus T$ respectively, we thus get that %in order for such a codeword $c$ to be erased, 
    at least one column-codeword of $\tilde{C_1}$ or one row-codeword of $\tilde{C_2}$ must be erased. That is, %if we call  the probability that some codeword $c=c^\textnormal{col}+c^\textnormal{row}$ with  is bounded by
% That is,
%\begin{align}\label{somerowcolfails}
   % \Pr_{x\sim p}[ \exists c&\in S_C(x) \textnormal{ with decomposition }S\times T ]\nonumber\\&\leq 2n^\alpha\cdot \max\left\{\Pr_{x\sim p}\left[\mathsf{decode}\big(\tilde{C_A},x\big)=0\right],\Pr_{x\sim p}\left[\mathsf{decode}\big(\tilde{C_B},x\big)=0 \right] \right\}.
%\end{align}
\begin{align}\label{somerowcolfails}
    \Pr_{\substack{C_1,C_2\\\bx\sim 1-\rho}}\Big[ \exists \bc&\in S_C(\bx) \textnormal{ of weight $w$ with supp($\bc_\textnormal{row})\subseteq S\times [n]$ and supp$(\bc_{\textnormal{col}})\subseteq [n]\times T$}\Big]\nonumber\\&\leq 2w\cdot \max\left\{\Pr_{\substack{C_1\\\bx\sim 1-\rho}}\left[\mathsf{decode}\big(\tilde{C_1},\bx\big)=0\right],\Pr_{\substack{C_2\\\bx\sim 1-\rho}}\left[\mathsf{decode}\big(\tilde{C_2},\bx\big)=0 \right] \right\}.
\end{align}
But whenever %$\Delta\geq \left(\frac{4}{\kappa\min\{ \rho_A(1-\rho_B),\rho_B(1-\rho_A)\}}\right)^{1/\tau}$
$n$ is large enough as a function of $\rho_1,\rho_2,\tau$, the code $\tilde{C_1}$ is a random linear code of rate
$$\frac{\textnormal{dim }\tilde{C_{1}}}{n-|S|}\leq \frac{\rho_1n}{n-\frac{n^{1-\tau}}{\kappa}}%\leq \rho_1+\frac{\rho_1n^{-\tau}}{\kappa}
\leq\rho_1+\frac{1}{2}\rho_2(1-\rho_1)
%\textnormal{dim }C_A^\perp-|S|%= \rho_A\Delta-\Delta^{1-\tau}
%\geq\rho_A\Delta-\frac{\min\{\rho_A(1-\rho_B),\rho_B(1-\rho_A)\}}{4}\Delta
$$
and $\tilde{C_2}$ is a random linear code of rate
$$\frac{\textnormal{dim }\tilde{C_2}}{n-|T|}\leq \frac{\rho_2n}{n-\frac{n^{1-\tau}}{\kappa}}\leq \rho_2+\frac{1}{2}\rho_1(1-\rho_2).
%\textnormal{dim }C_A^\perp-|S|%= \rho_A\Delta-\Delta^{1-\tau}
%\geq\rho_A\Delta-\frac{\min\{\rho_A(1-\rho_B),\rho_B(1-\rho_A)\}}{4}\Delta
$$
%where in both equation chains we used the fact that $\frac{1}{1-x}\leq 1+x.$ 
%This is true for every $w\leq n^{2-\tau}$ simultaneously. Taking a union bound over all weights $w\leq  n^{2-\tau}$ and applying Lemma \ref{randomcodescapacity}, we get that for all $n$ large enough
By Lemma \ref{randomcodescapacity}, both $\tilde{C_1}$ and $\tilde{C_2}$ can thus decode $(1-\rho_1-\rho_2+\rho_1\rho_2)$-noisy erasures with high probability. Thus for all $n$ large enough, equation (\ref{somerowcolfails}) becomes
\begin{align}\label{neweq}
    \Pr_{\substack{C_1,C_2\\\bx\sim 1-\rho}}\Big[ \exists \bc&\in S_C(\bx) \textnormal{ of weight $w$ with supp($\bc_\textnormal{row})\subseteq S\times [n]$ and supp$(\bc_{\textnormal{col}})\subseteq [n]\times T$}\Big]\nonumber\\
%&\leq \frac{4n^{1-\tau}}{\kappa} \cdot e^{-\frac{\min\left\{\rho_1(1-\rho_2),\rho_2(1-\rho_1)\right\}^2\left(n-\frac{n^{1-\tau}}{\kappa}\right)}{40}}\nonumber\\
%&\leq n \cdot e^{-\frac{\min\left\{\rho_1(1-\rho_2),\rho_2(1-\rho_1)\right\}^2n}{45}}.
&\leq e^{-bn}
%&\leq e^{-\frac{\varepsilon\Delta}{15}}%\left(2^{-D\left(\rho+\frac{\varepsilon}{2}||\rho\right)\Delta}+2^{-\frac{\varepsilon\Delta}{2}}\right)%\\
    %&\leq 2^{-\Omega(n)}.
\end{align}
for some constant $b=b(\rho_1,\rho_2,\tau)$.
%for $\varepsilon=\min\{\rho_A,\rho_B\}-\rho_A\rho_B-\Delta^{-\tau}$, where we used the facts that $\varepsilon\geq 0$  by our theorem's assumption on $\Delta$. 
Letting $E$ denote the event that $C$ is $\kappa$-product expanding and noting that there are at most $\binom{n}{w/\kappa n}\cdot \binom{n}{w/\kappa n}\leq 2^{\frac{2n^{1-\tau}}{\kappa} \log n}=2^{o(n)}$ choices for a pair $(S,T)$ with $|S|=|T|=\frac{w}{\kappa n}$, we get by equations \eqref{Cexpands}, \eqref{neweq} and the union bound that for all $n$ large enough, every $w\leq n^{2-\tau}$ satisfies
\begin{align*}
    \Pr_{\substack{C_1,C_2\\\bx\sim 1-\rho}}\left[ \exists \bc\in S_{C}(\bx) \textnormal{ with wt}(\bc)=w|E \right]%&\leq e^{-bn}\cdot2^{\frac{2n^{1-\tau}}{\kappa} \log n}\\
    &\leq e^{-\tilde{b}n}
    %&\leq n \cdot e^{-\frac{\min\left\{\rho_A(1-\rho_B),\rho_B(1-\rho_A)\right\}^2n}{45}}\cdot2^{\frac{2n^{1-\tau}}{\kappa} \log n}.   %\left(2^{-D\left(\rho+\frac{\varepsilon}{2}||\rho\right)\Delta}+2^{-\frac{\varepsilon\Delta}{2}}\right)%\\
    %&\leq 2^{-\Omega(n)}.
    %\\
   % &\leq \Delta \cdot e^{-\frac{\min\left\{\rho_A(1-\rho_B),\rho_B(1-\rho_A)\right\}^2\Delta}{50}}
\end{align*}
for some constant $\tilde{b}=\tilde{b}(\rho_1,\rho_2,\tau).$
Combining this with (\ref{Cexpands}) and a union bound over all weights $w\leq n^{2-\tau}$ gives our theorem statement.

\an{\textbf{ChatGPT comment.}
The current proof swaps the supports of the row and column terms in \eqref{somerowcolfails} and \eqref{neweq}; omits code randomness from those probabilities; and conditions on expansion while applying bounds for unconditioned random codes. It also drops the coefficient $5$ from its quoted expansion-failure bound. These are avoided by replacing the proof body by:
\begin{quote}
Let $E$ be the event that $C$ is $\kappa$-product expanding, with $\Pr(E^c)\leq e^{-b_0n}$ as in Theorem~\ref{hyperexpansionrandom}. On $E$, any nonzero $\bc\in C$ of weight $w\leq n^{2-\tau}$ has a decomposition $\bc=\bc_{\mathrm{col}}+\bc_{\mathrm{row}}$ whose nonzero columns are in $C_1^\perp$ and nonzero rows are in $C_2^\perp$. Let $T$ and $S$ be their column and row index sets. Then
\[
 |S|+|T|\leq\frac{w}{\kappa n},\qquad
 |S||T|\leq\frac{w^2}{4\kappa^2n^2}<w
\]
for sufficiently large $n$. Hence some nonzero entry lies outside $S\times T$. On that complement, the row and column terms cannot cancel. If $\bx$ covers $\bc$, it must therefore cover a nonzero word of $C_1^\perp$ punctured on $S$, in some column of $T$, or of $C_2^\perp$ punctured on $T$, in some row of $S$.
Put $M=\lfloor n^{1-\tau}/\kappa\rfloor$. For fixed $S,T$ of sizes at most $M=o(n)$, each punctured random code, conditional on its dimension, is a uniform subspace. Its rate is at most $\rho_i+o(1)$. At erasure probability $p_0+\eta$, the gap below the corresponding constituent threshold $1-\rho_i$ is at least $3g/4$. Lemma~\ref{randomcodescapacity}, after absorbing the puncturing and rounding terms, therefore bounds each punctured-code failure probability by $e^{-bn}$ for a constant $b>0$. A union bound over at most $2M$ relevant rows and columns gives $2Me^{-bn}$ for a fixed pair $S,T$.
There are at most $(M+1)^2n^{2M}=\exp(o(n))$ such support pairs. Taking the union bound over them and then adding $\Pr(E^c)$ proves the claimed $e^{-an}$ bound after decreasing $a>0$. All probabilities in these union bounds are over both the independent component codes and the independent erasure pattern; we bound $\Pr(E\cap\{\text{a small covered word exists}\})$ without conditioning the code distribution on $E$.
\end{quote}
This proves the needed statement for every weight in the specified interval simultaneously, rather than only one weight.}

\end{proof}

%Fix any $\tau\in(0,2)$ and $\rho_1,\rho_2\in(0,1)$. Then there exists some $n_0=n_0(\tau,\rho_1,\rho_2)$ such that for all $n>n_0$, the dual tensor code $C:=(C_1\otimes C_2)^\perp$ of two random linear codes $C_1,C_2\subseteq\F_2^n$ of rates $(1-\rho_1)$ and $(1-\rho_2)$ satisfies

\begin{theorem}\label{thmrandomduals}
Fix any $\rho_1,\rho_2\in(0,1)$ and $\nu\in(0,\frac{1}{3})$.% and let $C_1,C_2\subseteq\F_2^n$ be two random linear codes of rates $(1-\rho_1)$ and $(1-\rho_2)$ respectively. Then for $C:=(C_1\otimes C_2)^\perp$ and $\rho:=\rho_1+\rho_2-\rho_1\rho_2$, we have, for all $n$ larger than some constant $n_0=n_0(\rho_1,\rho_2,\nu)$,
Then there exists some $n_0=n_0(\nu,\rho_1,\rho_2)$ such that for all $n>n_0$, the dual tensor code $C:=(C_1\otimes C_2)^\perp$ of two independent uniform random linear codes $C_1,C_2\subseteq\F_2^n$ of rates $(1-\rho_1)$ and $(1-\rho_2)$ satisfies
\begin{align*}
\Pr_{\substack{{C_1,C_2}\\x\sim 1-\rho-n^{-\frac{1}{2}+\frac{3\nu}{2}}}}\Big[\mathsf{decode}(C,x)=0\Big]&\leq 4n^{-\frac{\nu}{2}}%,
\end{align*}
for $\rho:=\rho_1+\rho_2-\rho_1\rho_2$.
%where $a=a(\nu,\rho_1,\rho_2)$ is a positive constant  that depends on $\nu,\rho_1$ and $\rho_2$ but not on $n$.
\end{theorem}

% old
% \begin{theorem}\label{thmrandomduals}
% Fix any $\rho_1,\rho_2\in(0,1)$ and $\nu\in(0,\frac{1}{3})$.% and let $C_1,C_2\subseteq\F_2^n$ be two random linear codes of rates $(1-\rho_1)$ and $(1-\rho_2)$ respectively. Then for $C:=(C_1\otimes C_2)^\perp$ and $\rho:=\rho_1+\rho_2-\rho_1\rho_2$, we have, for all $n$ larger than some constant $n_0=n_0(\rho_1,\rho_2,\nu)$,
% Then there exists some $N_0=N_0(\nu,\rho_1,\rho_2)$ such that for all $N>N_0$, the dual tensor code $C:=(C_1\otimes C_2)^\perp$ of two random linear codes $C_1,C_2\subseteq\F_2^N$ of rates $(1-\rho_1)$ and $(1-\rho_2)$ satisfies
% \begin{align*}
% \Pr_{\bx\sim 1-\rho-N^{-\frac{1}{2}+\frac{3\nu}{2}}}\Big[\mathsf{decode}(C,\bx)=0\Big]&\leq 4N^{-\frac{\nu}{2}}%,
% \end{align*}
% for $\rho:=\rho_1+\rho_2-\rho_1\rho_2$.
% %where $a=a(\nu,\rho_1,\rho_2)$ is a positive constant  that depends on $\nu,\rho_1$ and $\rho_2$ but not on $n$.
%\end{theorem}

\begin{proof}
%{\color{red}In this proof, it would be nice to preserve the following intuition/keep the following chain of implications as clear as possible, because I think it is useful for the reader: $C_1$ and $C_2$ achieve capacity, so $C_1\otimes C_2$ achieves capacity (on a channel of different noise). And because of that $(C_1\otimes C_2)^\perp$ achieves capacity (on a channel of yet another noise).} {\color{\markcolor} I did my best.  Feel free to edit.}

We first bound the decoding error probabilities of $C_1$ and $C_2$ near capacity: by Lemma~\ref{randomcodescapacity}, for $\gamma={n^{-\frac{1-\nu}{2}}}$ and any $n$ large enough, we have
    \begin{align*}
  \Pr_{\substack{C_1\subseteq\F_2^n\\\bx\sim \rho_1-\frac{\gamma}{2}}}\big[\mathsf{decode}(C_1,\bx)=0  \big]&\leq %2e^{-\frac{\varepsilon^2 \Delta}{48}}        \leq 
        e^{-bn^{\nu}},
    \end{align*}
    \begin{align*}
        \Pr_{\substack{C_2\subseteq\F_2^n\\\bx\sim \rho_2-\frac{\gamma}{2}}}\big[\mathsf{decode}(C_2,\bx)=0  \big]&\leq e^{-bn^{\nu}}
    \end{align*}
    for some constant $b=b(\nu).$
Thus the code $C_1\otimes C_2$ has a small decoding error probability and achieves capacity. More specifically, since
$$\rho-\gamma\leq \tilde{\rho}_1+\tilde{\rho}_2-\tilde{\rho}_1\tilde{\rho}_2$$
for $\tilde{\rho}_1=\rho_1-\frac{\gamma}{2}$ and $\tilde{\rho}_2=\rho_2-\frac{\gamma}{2}$, Theorem \ref{tensorcodescapacity} implies that
    \begin{align}\label{tensorcap}
        \E_{\substack{C_1,C_2\\\bx\sim \rho-\gamma}}\Big[\textnormal{dim }S_{C^\perp}(\bx)\Big]& \leq  2n^3e^{-bn^{\nu}}.%\\
       % &\leq e^{-\sqrt{\Delta}}.%n^4e^{-\Omega(\delta^2 n)}. 
        %o(1).
    \end{align}

We will now show that \eqref{tensorcap} implies that the code $C=(C_1\otimes C_2)^\perp$ has a small decoding error probability and achieves capacity. Let $\bA\in\F_2^{(n^2-\textnormal{dim }C)\times n^2}$ be a generator matrix of $C^\perp=C_1\otimes C_2.$ 
    Note that %since $C_A\otimes C_B$ achieves capacity, we have (by Lemma \ref{equivalentformulationscoveredsubcode})
    by equation \eqref{tensorcap} and Lemma \ref{equivalentformulationscoveredsubcode}, we have
    \begin{align*}
        \E_{\substack{C_1,C_2\\\bx\sim \rho-\gamma}}[\textnormal{rank }\bA_{\bar{\bx}}]\geq\textnormal{dim }C^\perp-2n^3e^{-bn^{\nu}},%e^{-\sqrt{\Delta}},%4n^4e^{-\frac{\delta^2n}{48}},
    \end{align*}
    %where we denote by $G_{\bar{x}}$ the restriction of $G$ to the rows $i\in[n^2]$ where $\bar{x}_i=1$ (or equivalently, where $x_i=0$). 
    %for $p:=p_1+p_2-p_1p_2$, 
    where for any vector $\bz\in\F_2^n$, we denote by $\bA_{\bz}$ the restriction of the matrix $\bA$ to the columns $i\in[n]$ where $\bz_i=1$. Thus
    \begin{align}\label{reformulate}
        \E_{\substack{C_1,C_2\\\bx\sim 1-\rho+\gamma}}[\textnormal{rank }\bA_{\bx}]\geq\textnormal{dim }C^\perp-2n^3e^{-bn^{\nu}}.%e^{-\sqrt{\Delta}}.%4n^4e^{-\frac{\delta^2n}{48}}\
    \end{align}
But by definition, $\bA$ is also the parity-check matrix of $C$, so Lemma \ref{equivalentformulationscoveredsubcode} and equation \eqref{reformulate} give
\begin{align*}
    \E_{\substack{C_1,C_2\\\bx\sim 1-\rho+\gamma}}[\textnormal{dim }S_{C}(\bx)]%&=\E_{x\sim 1-\rho+\gamma}\Big[\textnormal{wt}(x)-\textnormal{rank } A^\intercal_{\textnormal{cols }x}\Big]\nonumber\\
    &= \E_{\substack{C_1,C_2\\\bx\sim 1-\rho+\gamma}}[\textnormal{wt}(\bx)-\textnormal{rank } \bA_{\bx}\Big]\nonumber\\
    &\leq \gamma n^2+2n^3e^{-bn^{\nu}}\nonumber\\%e^{-\sqrt{\Delta}}\nonumber\\%4n^4e^{-\frac{\delta^2n}{48}}.
%&\leq 2\Delta^{3/2+\nu/2}.\\
&= n^{\frac{3+\nu}{2}}+2n^3e^{-bn^{\nu}}.
\end{align*}
%where for any vector $z\in\F_2^N$, we denote by $A_{\textnormal{cols }z}$ the restriction of the matrix $A$ to the columns $i\in[N]$ where $z_i=1$.
In particular, by Markov's inequality and monotonicity of the erasure channel, for all $n$ large enough we have
\begin{align}\label{bddimsubcode}
 \Pr_{\substack{C_1,C_2\\\bx\sim 1-\rho}}\left[\textnormal{dim }S_{C}(\bx)\geq n^{\frac{3}{2}+\nu}\right]\leq 2n^{-\frac{\nu}{2}}.
\end{align}
We have thus established that when the noise parameter is slightly above $1-\rho$, the covered subcode of $C$ has small dimension with high probability. But by Lemma \ref{nosmallcodeword}, every codeword in this covered subcode must have large weight. Intuitively, this means that if we randomly uncover a few erasures, we will with high probability uncover at least one erasure in each of those covered codewords. Formally, 
 % All that remains to show is now that decreasing the noise a little bit brings this expected dimension down to $o(1).$ We think of sampling a $(1-p-\delta)$-noisy string $\tilde{x}$ by first sampling a $(1-p+\delta )$-noisy string $x$ and then repeatedly removing batches of at most $ n^{\alpha}$ erasures. Formally, we %Define the bad events
%\begin{align*}
%    B(y,C)&:=\begin{cases}
%1 & \text{if dim $S_C(y)>\delta n^2$ },\\
%0 & \text{otherwise},
%\end{cases}
%\end{align*}
%\begin{align*}
%    \tilde{B}(y,C)&:=\begin{cases}
%1 & \text{if $\exists c\in S_{C}(y):\textnormal{wt}(c)\leq n^{2-\alpha}$ },\\
%0 & \text{otherwise},
%\end{cases}
%\end{align*}
%and recall that by equation \eqref{bddimsubcode} and Lemma \ref{nosmallcodeword} we have
%\begin{align*}
   % \Pr_{y,C}\left[B(y,C)=\tilde{B}(y,C)=0\right]=1-o(1).
%\end{align*}
let $\bz^1,\dotsc,\bz^{\lfloor n^{\frac{3}{2}+\nu}\rfloor}\in\F_2^{n^2}$ be independent $n^{-\left(2-\frac{\nu}{2}\right)}$-noisy variables and define for each $i\in\{1,\dotsc,\lfloor n^{\frac{3}{2}+\nu}\rfloor\}$ the erasure pattern
\begin{align*}
    \bx^i:=\bx\setminus\left( \bz^1\vee\dotsc\vee \bz^i\right).
\end{align*}
Consider the event $E=E(\bx)$ defined by
\begin{align*}
   E(\bx)&:=\begin{cases}
1 & \text{if $\textnormal{dim }S_{C}(\bx)\leq n^{\frac{3}{2}+\nu}$ and $\textnormal{wt}(\bc)\geq {n^{2-\frac{\nu}{4}}}\textnormal{ for all }\bc\in S_C(\bx)\setminus\bzero$},\\
0 & \text{otherwise}.
\end{cases}
\end{align*}
%Let $E$ be the event that $\textnormal{dim }S_{C}(x)\geq n^{\frac{3}{4}}\log^2 n$ and that $\big|\textnormal{supp}(S_C(x))\big|\leq n^{0.9}$. 
Note that by equation \eqref{bddimsubcode} and Lemma \ref{nosmallcodeword}% (where for the latter, we take a union bound over at most $n^2$ possible weights)
, for all $n$ large enough we have
\begin{align}\label{Ehappens}
\Pr_{\substack{C_1,C_2\\\bx\sim 1-\rho}}[E(x)=1]&\geq 1-3n^{-\frac{\nu}{2}}.%&\geq 1-n^{-\frac{\nu}{2}}-n^2e^{-\frac{n^{\nu}}{40}}-5\cdot2^{-\frac{\rho_A\rho_B}{8}n^2}\nonumber\\
%&\quad-16\cdot 2^{-\frac{\rho_A\rho_B}{8}n}+n \cdot e^{-\frac{\min\left\{\rho_A(1-\rho_B),\rho_B(1-\rho_A)\right\}^2n}{45}}\cdot2^{\frac{2n^{1-\nu/4}}{\kappa} \log n}.
\end{align}

\an{\textbf{ChatGPT comment.}
As written, $E$ requires every word of $S_C(\bx)$, including zero, to have positive weight, so $E$ is impossible. Replace its definition by:
\[
 E(C,\bx)=\left\{\dim S_C(\bx)\leq n^{3/2+\nu},\quad
  |\bc|\geq n^{2-\nu/4}\ \text{for every }\bc\in S_C(\bx)\setminus\{0\}\right\}.
\]
Replace the next probability bound by
\[
 \Pr_{C_1,C_2,\,\bx\sim(1-\rho+\gamma)_{n^2}}[E(C,\bx)]
 \geq1-2n^{-\nu/2}-e^{-an}\geq1-3n^{-\nu/2}.
\]
This uses updated result with $\tau=\nu/4$, since $\gamma\to0$ is eventually smaller than its fixed $\eta$. Without strengthened noise and weight range, the cited lemma does not imply this event. Also replace ``slightly above $1-p$'' by ``slightly above $1-\rho$'' and change the matrix-restriction sentence to $\bz\in\F_2^{n^2}$, $i\in[n^2]$.}
Let $E_i(\bx)$ be the event that dim $S_C(\bx^i)>0$. Since $\big|\textnormal{supp }S_C(\bx^i)\big|\geq n^{2-\frac{\nu}{4}}$ whenever $E(\bx)=E_i(\bx)=1$, we have 
\begin{align*}%\label{goaldimwin}
    \Pr_{\substack{C_1,C_2\\\bx\sim 1-\rho \\\bz^1,\dotsc,\bz^{i+1}\sim n^{-\left(2-\frac{\nu}{2}\right)}}}\Big[\textnormal{dim }S_C\left(\bx^{i+1}\right)<\textnormal{dim }S_C(\bx^i)\big|E,E_i\Big]&\geq 1-\left(1-n^{-\left(2-\frac{\nu}{2}\right)}\right)^{n^{2-\frac{\nu}{4}}}\\
&\geq 1-e^{-n^{\frac{\nu}{4}}},
\end{align*}
where in the last line we used the inequality $1-a\leq e^{-a}$.
So with high probability, every $S_C\left(\bx^i\right)$ of nonzero dimension has smaller dimension than its predecessor $S_C\left(\bx^{i-1}\right)$. We will thus exhaust the original subcode cover's dimension after at most $n^{\frac{3}{2}+\nu}$ steps. Formally, by equation \eqref{Ehappens} and the union bound, we have that for all $n$ large enough,
\begin{align*}
 \Pr_{\substack{C_1,C_2\\\bx\sim 1-\rho \\\bz^1\dotsc,\bz^{\lfloor n^{\frac{3}{2}+\nu}\rfloor}\sim n^{-\left(2-\frac{\nu}{2}\right)}}}\Big[\textnormal{dim }S_C\left(\bx^{\lfloor n^{\frac{3}{2}+\nu}\rfloor}\right)>0\Big]&\leq \Pr%_{x\sim \rho}
 [E(\bx)=0]\\
&+\Pr%_{\substack{x\sim 1-\rho+\varepsilon \\\{z^i\}\sim \Delta^{-\left(2-\frac{\nu}{2}\right)}}}
\Big[\exists i:\textnormal{dim }S_C\left(\bx^{i+1}\right)=\textnormal{dim }S_C(\bx^i)>0\big|E\Big]\\
&\leq 4 n^{-\frac{\nu}{2}}.
\end{align*}
%for $\varepsilon=n^{-\frac{\nu}{2}}+n^2e^{-\frac{n^{\nu}}{40}}+5\cdot2^{-\frac{\rho_A\rho_B}{8}n^2}+16\cdot 2^{-\frac{\rho_A\rho_B}{8}n}+n \cdot e^{-\frac{\min\left\{\rho_A(1-\rho_B),\rho_B(1-\rho_A)\right\}^2n}{45}}\cdot2^{\frac{2n^{1-\nu/4}}{\kappa} \log n}+n^2e^{-n^{\frac{\nu}{4}}}$.
Now the erasure pattern $\tilde{\bx}=\bx\setminus\left(\bz^1\vee\dotsc\vee \bz^{\lfloor n^{\frac{3}{2}+\nu}\rfloor}\right)$ is $\mu$-noisy for 

\noindent$\mu=(1-\rho)\left(1-n^{-\left(2-\frac{\nu}{2}\right)}\right)^{\lfloor n^{\frac{3}{2}+\nu}\rfloor}\geq (1-\rho)\left(1-n^{-\frac{1}{2}+\frac{3\nu}{2}}\right)\geq 1-\rho-n^{-\frac{1}{2}+\frac{3\nu}{2}}$, so our previous equation implies that
\begin{align*}
\Pr_{\substack{C_1,C_2\\\bx\sim 1-\rho-n^{-\frac{1}{2}+\frac{3\nu}{2}}}}\Big[\mathsf{decode}(C,\bx)=0\Big]&\leq %\Pr_{\substack{x\sim 1-p+\delta \\\{z^i\}\sim n^{-1.8}}}\Big[\textnormal{dim }S_C\left(x^{n^{3/2}\log^2 n}\right)>0\Big]+o(1)\\
%&=o(1).
%\frac{1}{\Delta}+\frac{1}{\sqrt{\ln \Delta}}+o(1).
%\varepsilon.
4n^{-\frac{\nu}{2}}.
\end{align*}

\an{\textbf{ChatGPT comment.}
Use an integer number of rounds by replacing every round count $n^{3/2+\nu}$ in the definition of the batches and their indices by $K:=\lfloor n^{3/2+\nu}\rfloor$; keep the same real bound in the good event. Since the dimension is an integer, that event bounds it by $K$. Add:
\begin{quote}
Sample the batches independently of the component codes and the initial erasure pattern, and put $\bx^0=\bx$. Conditional on any history with $E(C,\bx)$ and $S_C(\bx^i)\ne\{0\}$, the next batch hits the support of that subcode except with probability at most $e^{-n^{\nu/4}}$. Hitting that support strictly reduces the dimension, so a union bound over $K$ rounds suffices.
\end{quote}
In the final sentence replace ``can be rewritten as'' by ``implies the following bound by monotonicity in the erasure probability'', since the calculated batch distribution has parameter $\mu$ at least the target parameter, rather than equal to it. The stated $4n^{-\nu/2}$ conclusion then follows.}

%Now with high probability, the erasure pattern $\tilde{x}=x\setminus\left(z^1\vee\dotsc\vee z^{n^{3/2}\log^2 n}\right)$ has at least $(1-p)n-n^{1.74}$ erasures, whereas a $(1-p-n^{-1/4})$-noisy string has weight $<(1-p)n-n^{1.74}$ with high probability. Since both distributions are uniformly distributed on any fixed weight class, we have
%\begin{align*}
%\Pr_{x\sim 1-p-n^{-1/4}}\Big[\textnormal{dim }S_C\left(x\right)>0\Big]&\leq \Pr_{\substack{x\sim 1-p+\delta \\\{z^i\}\sim n^{-1.8}}}\Big[\textnormal{dim }S_C\left(x^{n^{3/2}\log^2 n}\right)>0\Big]+o(1)\\
%&=o(1).
%\end{align*}

\end{proof}

% Theorem~\ref{thmrandomduals} implies capacity for the dual tensor-code
% family in Definition~\ref{def:tensorcodefamilies}. Indeed, its rate is
% $1-(1-\rho_1)(1-\rho_2)$, up to integer rounding, while its erasure threshold
% approaches $(1-\rho_1)(1-\rho_2)$ from below and its ensemble-average error
% tends to zero. Choosing the component rates so that their product lies
% strictly between a prescribed erasure probability and that probability
% plus the desired rate gap gives \eqref{familycapacity} at arbitrarily
% large lengths.

\section{Capacity-achieving quantum Tanner codes}\label{section:quantumTannercapacity}

In this section, we prove that the quantum Tanner code construction described in Section~\ref{sec:QTC} achieves capacity on the erasure channel with a positive error exponent. 

\subsection{Theorems~\ref{thm:LZ2023} and \ref{thm:decoding}}
First, we give some comments on the proofs of Theorems~\ref{thm:LZ2023} and \ref{thm:decoding} and the fact that the generalisation to four constituent inner codes $C_A,C_B,D_A,D_B$ instead of two leaves the proofs of \cite{LZ2023} unchanged. 

Consider Theorem~\ref{thm:LZ2023} on the minimum distance of the quantum Tanner code. We consider a codeword $\bx$ of one of the two codes, say $C_X$, and want to prove that if its weight is too small, then it must belong to the row-space of $\bH_Z$, namely $C_Z^\perp$. The codeword $\bx$ may be thought of as subset of squares, and since the code $C_X$ is defined by equations that are supported by the local views at vertices of $V_{00}$ and $V_{11}$, we may momentarily forget the vertices of $V_{01}$ and $V_{10}$ so that squares become incident to only two vertices, thus defining the edges of a bipartite graph $\mathcal{G}$. If the component Cayley graphs $\Cay(A,G)$ and $\Cay(G,B)$ are chosen to be Ramanujan graphs, then the graph $\mathcal{G}$ has degree $\Delta^2$ and second eigenvalue not more than $4\Delta$~\cite[Lemma 4]{LZ22}. Standard expansion of graphs arguments tell us that any subgraph of $\mathcal{G}$ that has average degree more than $4\Delta$ must be linear in the number of edges of $\mathcal{G}$, i.e. the length of $C_X$. Conversely, if the weight of our codeword $\bx$ is chosen to be smaller than this quantity, it must have an average degree that scales like $\Delta$. 

%%Gilles: strongly disagreed with that ChatGPT comment. Ignored.
%{\textbf{ChatGPT comment.}
%A subgraph having average degree merely greater than the spectral bound does not by itself give a uniform linear-size conclusion without a quantitative excess. Replace the two sentences starting ``Standard expansion'' by:
%\begin{quote}
%For vertex subsets $S\subseteq V_{00}$ and $T\subseteq V_{11}$, expander mixing gives
%\[
% |E(S,T)|\leq\frac{\Delta^2|S||T|}{|G|}+4\Delta\sqrt{|S||T|}.
%\]
%In particular, for a supported subgraph on these vertices its average degree is at most $\Delta^2(|S|+|T|)/(2|G|)+4\Delta$. Thus a small active vertex set has average degree $O(\Delta)$, which is the estimate used to distinguish ordinary and exceptional local views in the distance proof.
%\end{quote}
%%Gilles: agreed with that below, followed instructions.

%The distance argument starts with $\bx\in C_X$ and aims to prove $\bx\in\operatorname{row}\bH_Z=C_Z^\perp$. In the next paragraph replace ``vertices $vi$'' by ``vertices $v$'', ``row codewords of $C_A^\perp$ and ... column codewords of $C_B^\perp$'' by ``column codewords of $C_A^\perp$ and row codewords of $C_B^\perp$'', and interchange rows/columns in the later sentence pairing $C_A^\perp$ with $D_B^\perp$. In the last sentence replace ``i.e. should be $\omega(\sqrt\Delta)$'' by ``which holds for sufficiently large $\Delta$ after fixing $\delta$ and $\kappa$''; the asymptotic growth is part of the parameter selection, not the limit in global blocklength.}

After arriving at this first conclusion, we consider how typically $\bx$ intersects non-trivially local views of vertices of $V_{00}\cup V_{11}$ and invoke Theorem~\ref{hyperexpansionrandom} to claim that pairs 
$C_A,C_B$ and $D_A,D_B$ have some fixed robustness (or product expansion) parameter $\kappa$. This means that typical local views of vertices $v\in V_{00}$ incident to $\bx$ decompose into a sum of a constant number of column codewords of $C_A^\perp$ and a constant number of row codewords of $C_B^\perp$ (for $v\in V_{11}$ $C_A$ and $C_B$ are replaced by $D_A$ and $D_B$). We then look at
the typical local views of vertices incident to $\bx$, but for the {\em other} vertex types $V_{01}$ and $V_{10}$. 
Using the local structure of the square complex and the expansion properties of $\Cay(A,G)$ and $\Cay(G,B)$ one obtains that the local views for these vertices are typically very different.
The sparse sets of rows and columns that support the local views for $V_{00}\cup V_{11}$ must cluster around the local views for $V_{01}\cup V_{10}$. We obtain that some local views at $v\in
V_{01}$ must be made up of large number of rows that are almost all close to codewords of $C_A^\perp$ and they also consist of a large number of columns that are almost all close to codewords of $D_B^\perp$. Here 'large' means at least a quantity that is governed by the smallest minimum distance of the two constituent codes $C_A^\perp$ and $D_B^\perp$. 
(The vertex $v$ may possibly need to belong to $V_{10}$ instead of $V_{01}$, in which case the codes
$C_A^\perp$ and $D_B^\perp$ need to be replaced by $D_A^\perp$ and $C_B^\perp$).
Robustness is then again invoked to claim that such a pattern must be close to a codeword of the tensor code $C_A^\perp\otimes D_B^\perp$. But this is exactly a parity-check equation of $C_Z$, which we may therefore add to $\bx$ to obtain a codeword of $C_X$ of smaller weight. Iterating gives us that $\bx$ must belong to $C_Z^\perp$ which is the desired behaviour.

The point of this sketch is to stress that the proof only needs to know lower bounds $\delta$
and $\kappa$ on the relative minimum distances and the robustness of the relevant constituent codes and pairs of constituent codes. The rest consists of the combinatorics of expansion and averaging arguments. Once $\delta$ and $\kappa$ are fixed by the rates $\rho_A,\rho_B,1-\rho_A,1-\rho_B,\varepsilon,1-\varepsilon$ of the constituent codes, a sufficiently large $\Delta$ should be chosen so that expansion can apply, notably $\delta\Delta$ and $\kappa\Delta$ should exceed significantly the second eigenvalue of the graphs, $\Cay(A,G)$ and $\Cay(G,B)$, 
which holds for sufficiently large $\Delta$ after fixing $\delta$ and $\kappa$.
%i.e. should be $\omega(\sqrt{\Delta})$.

For Theorem~\ref{thm:decoding} the proof of \cite{LZ2023} is more involved but follows the same principle. We are given the syndrome of an error vector $\be_Z$, and to decode it we apply a local decoding algorithm at every vertex of $V_{00}\cup V_{11}$. If all the local decoders agree on every coordinate, then we have global candidate for the error vector. This is unlikely to happen though, there will be coordinates for which the two local views to which it belongs, in $V_{00}$ and $V_{11}$, disagree. The set of positions for which the local decoders disagree is called the {\em mismatch} vector. It is equal to a sum of dual tensor codewords supported by local views of vertices of $V_{00}\cup V_{11}$. Robustness ensures that every dual tensor codeword decomposes as a sum of row codewords and column codewords. Every row codeword and every column codeword can be shared with a neighbouring vertex whose local view shares this very row or column. Therefore the mismatch vector can also be read locally on the local views of vertices of $V_{01}\cup V_{10}$.
Analysing the local structure of the mismatch vector leads us to conclude that it must be possible to add a dual tensor codeword on the local view of one of the four types of vertices that decreases the mismatch weight. Iterating eventually reduces the mismatch to zero, after which we have a candidate for the error vector. Again, the proof consists of leveraging expansion arguments, with the only coding requirements being that local dual tensor codewords can be faithfully decomposed as a sum of row codewords and column codewords (robustness) and that everyone of those codewords has a sufficiently large minimum weight (distance).

\subsection{Analysing the behaviour on the erasure channel}
For convenience, we briefly recall our construction: fix any parameters $\rho\in(0,\frac{1}{2})$ and any small $\varepsilon$. We consider a quadripartite Cayley square complex with vertex set $V_{00}\cup V_{11}\cup V_{01}\cup V_{10}$ and generators $A,B$ of sizes $|A|=|B|=\Delta$. We place the qubits of our code on the squares of the complex. We let the vertices of $V_{00}$ and $V_{11}$ support the $X$-check equations and we let the vertices of $V_{10}$ and $V_{01}$ support the $Z$-check equations. By construction, each check equation is thus supported on a set of $\Delta^2$ qubits that can naturally be indexed by $A\times B$. We take the $V_{00}$ check equations to come from $C_A\otimes C_B$, the $V_{11}$ check equations to come from $D_A\otimes D_B$, the $V_{01}$ check equations to come from $C_A^\perp\otimes D_B^\perp$ and the $V_{10}$ check equations to come from $D_A^\perp\otimes C_B^\perp$, where $C_A,C_B,D_A,D_B$ are linear codes of length $\Delta$ and rate
    \begin{align*}
        \textnormal{rate}(C_A)=1-\rho-\frac{\varepsilon}{3},
    \end{align*}
    \begin{align*}
        \textnormal{rate}(C_B)=\frac{\rho+\frac{\varepsilon}{6}}{1-\rho-\frac{\varepsilon}{3}},
    \end{align*}
    \begin{align*}
        \textnormal{rate}(D_A)=1-\frac{\varepsilon}{4},
    \end{align*}
    \begin{align*}
        \textnormal{rate}(D_B)=\frac{\varepsilon}{4},
    \end{align*}
    for some fixed $\varepsilon\in(0,1-2\rho)$.
 As we established in Section~\ref{sec:QTC}, this code is a valid CSS code. We compute
 $$\textnormal{rate}(C_A\otimes C_B)=\left(1-\rho-\frac{\varepsilon}{3}\right)\frac{\rho+\frac{\varepsilon}{6}}{1-\rho-\frac{\varepsilon}{3}}= \rho+\frac{\varepsilon}{6},$$ 
 $$\textnormal{rate}(C_A^\perp\otimes D_B^\perp)=(\rho+\frac{\varepsilon}{3})\left(1-\frac{\varepsilon}{4}\right),$$ 
 $$\textnormal{rate}(D_A\otimes D_B)\leq\frac{\varepsilon}{4},$$ 
 $$\textnormal{rate}(D_A^\perp\otimes C_B^\perp)\leq\frac{\varepsilon}{4}.$$ 
 In particular, we have
 \begin{align*}%\label{goodrate}
     \textnormal{rate}(C_{\rho,\varepsilon,\Delta,n})&\geq %1-\left(p+\frac{\varepsilon}{4}\right)-\left(p+\frac{\varepsilon}{2}\right)+\frac{\varepsilon}{4}=
     1-2\rho-\varepsilon
 \end{align*}
 but also
  \begin{align*}%\label{goodrateX}
      \textnormal{rate}(C_A\otimes C_B)\geq \rho+\frac{\varepsilon}{12},
  \end{align*} 
  \begin{align*}%\label{goodrateZ}
      \textnormal{rate}(C_A^\perp\otimes D_B^\perp)\geq \rho+\frac{\varepsilon}{12}.
  \end{align*}
 Thus we have obtained a quantum Tanner code of rate arbitrarily close to $1-2\rho$ for which the $X$ and $Z$ checks associated with the vertices in $V_{00}$ and $V_{01}$ come from codes of rates larger than $\rho$. By Theorem \ref{thmrandomduals}, these two codes typically succeed in decoding $\rho$-noisy erasures; we should thus be able to decode $\be_Z$ almost entirely by using the $X$-checks associated with the vertices of $V_{00}$, and we should be able to decode $\be_X$ almost entirely by using the $Z$-checks associated with the vertices of $V_{01}$.

More formally, by Theorem \ref{hyperexpansionrandom}, there exists constants $\delta,\kappa$ depending only on $\rho$ and $\varepsilon$ such that we can choose $C_A,C_B,D_A,D_B$ and their duals to have minimum distance $\delta \Delta$ and any pair (except for those comprised of a component code and its dual) to be $\kappa$-robust. (For $\Delta$ large enough, a random choice of $C_A,C_B,D_A,D_B$ satisfies these requirements with high probability.) Furthermore, by Theorem \ref{thmrandomduals}, as long as $\Delta$ is large enough in terms of $\rho,\varepsilon$, we can additionally choose our component codes so that $C_A\otimes C_B$ and $C_A^\perp\otimes D_B^\perp$ satisfy

 \begin{align}\label{firstgood}
\Pr_{\substack{x\sim \rho+\frac{\varepsilon}{15}}}\Big[\mathsf{decode}\left((C_A\otimes C_B)^\perp,x\right)=0\Big]&\leq \Delta^{-\frac{1}{10}},%,
\end{align}
 \begin{align}\label{secondgood}
\Pr_{\substack{x\sim \rho+\frac{\varepsilon}{15}}}\Big[\mathsf{decode}\left(\left(C_A^\perp\otimes D_B^\perp\right)^\perp,x\right)=0\Big]&\leq \Delta^{-\frac{1}{10}}.%,
\end{align}
(Again, a random choice of component codes will satisfy these requirements with high probability.) Denote by $C_{\rho,\varepsilon,\Delta,n}$ any such construction of Tanner codes, i.e. any construction above with $C_A,C_B,D_A,D_B$ and their duals  having minimal distance at least $\delta \Delta$, robustness at least $\kappa$, and satisfying equations \eqref{firstgood} and \eqref{secondgood}. In our next theorem, which is a generalization of Theorem \ref{thm:qTannerIntro}, we show that $C_{\rho,\varepsilon,\Delta,n}$ achieves capacity on the quantum erasure channel.

\an{\textbf{ChatGPT comment.}
An arbitrary $\varepsilon>0$ can also make the prescribed component rate exceed one. Replace the parameter choice and rate calculations, starting ``For convenience'', by the following block (retaining the construction description):
\begin{quote}
Fix $\rho\in(0,1/2)$ and a desired rate gap $\varepsilon>0$. Choose
\[
 0<\eta<\min\{\varepsilon/2,(1-2\rho)/2,1/2\}.
\]
Our ideal local rates are
\[
 r_A=1-\rho-\eta/3,\qquad
 r_B=\frac{\rho+\eta/6}{1-\rho-\eta/3},\qquad
 s_A=1-\eta/4,\qquad s_B=\eta/4.
\]
All four rates lie strictly between zero and one. Choose independent uniform local codes with dimensions obtained by rounding these rates times $\Delta$ to integers. Before rounding,
\[
 r_Ar_B=\rho+\eta/6,\qquad
 (1-r_A)(1-s_B)=(\rho+\eta/3)(1-\eta/4)\geq\rho+\eta/6,
\]
while $s_As_B\leq\eta/4$ and $(1-s_A)(1-r_B)\leq\eta/4$. The sum of the four local check-code rates is at most $2\rho+\eta$. For sufficiently large $\Delta$, the $O(1/\Delta)$ rounding errors preserve
\[
 \operatorname{rate}(\mathcal C)\geq1-2\rho-\varepsilon,
 \quad\operatorname{rate}(C_A\otimes C_B)\geq\rho+\eta/12,
 \quad\operatorname{rate}(C_A^\perp\otimes D_B^\perp)\geq\rho+\eta/12.
\]
\end{quote}
Keep the corresponding equation labels on the last three inequalities. Use $\eta$ for this local slack throughout the proof below; the final rate gap remains $\varepsilon$. Write the family as $C_{\rho,\varepsilon,\Delta,N}$, using global length $N$ consistently. }

\begin{theorem}\label{thm:TannerCapacity}
    Consider any $\rho\in(0,\frac{1}{2})$ and any $\varepsilon\in(0,1-2\rho)$. %Then there exists some constant $\Delta_0=\Delta_0(R,\varepsilon)$ such that for all $\Delta\geq \Delta_0$, the quantum Tanner code $C$ constructed in ?? satisfies
    Then there exists positive constants %$\Delta=\Delta(\rho,\varepsilon),a=a(\rho,\varepsilon)$ and $N_0=N_0(\rho,\varepsilon)$ 
    $a,\Delta,N_0$ (i.e. depending only on $\rho$ and $\varepsilon$) such that for all $N>N_0$, the quantum Tanner code $ C_{\rho,\varepsilon,\Delta,N}$ has rate at least $1-2\rho-\varepsilon$  and can decode random quantum erasures of rate $\rho$ with probability%$C$ constructed in ?? satisfies
    \begin{align*}
        \Pr_{\beps\sim \rho}[\mathsf{decode}(C_{\rho,\varepsilon,\Delta,N},\beps)=1]\geq 1- 2^{-aN}
    \end{align*}
    and decoding time $\leq \frac{N}{a}.$
\end{theorem}

\begin{proof}
We are given a subset $\beps\subseteq[N]$ of erased coordinates and the syndromes $\bH_X\be_Z^\intercal$ and $\bH_Z\be_X^\intercal$ of two vectors $\be_X,\be_Z\in\F_2^N$ that were sampled uniformly randomly on $\beps$ and are identically $0$ everywhere else. Our goal is to output the cosets $\be_X+C^\perp_X$ and $\be_Z+C_Z^\perp.$ 

Our decoding algorithm is quite simple: first, go through every vertex $v\in V_{00}$ and attempt to decode the vector $\be_v$ from the erasure set $\beps$ and the syndrome $\bH_v\be_v^\intercal$, where $\be_v$ denotes the restriction of the error vector $\be_Z$ to the local view $Q(v)$ and $\bH_v$ denotes the restriction of the parity-check matrix $\bH_{X}$ to the set of constraints imposed by the vertex $v.$\footnote{Algorithmically, this is done by performing Gaussian elimination to solve the system of equations $\bm{A}\bm{x}=\bm{b}$ for $\bm{A}$ the restriction of $\bH_v$ to the columns indexed by $\beps$, $\bm{x}$ the restriction of $\be_v$ to the coordinates indexed by $\beps$, and $\bm{b}$ the syndrome $\bH_v\be_v^\intercal$ and accepting the solution if and only if it is unique.\label{footnotealgo}}

%That is, letting $S\subseteq V_{00}$ be the set of vertices $v\in V_{00}$ for which $\be_v$ can be uniquely determined by its syndrome $\bH_v\be_v^\intercal$ and defining 
%$$T:=\mathop{\cup}_{v\in S}Q(v),$$ 
%let
%\begin{align*}
 %   [\bu_X]_{i}:=\begin{cases}
%[\be_X]_{i} & \text{if $i\in T$,}\\
%0 & \text{otherwise.}
%\end{cases}
%\end{align*}

Then, proceed in exactly the same way for the $X$-component of the error: go through every vertex $v\in V_{01}$ and attempt to decode the vector $\be_v$ from the erasure set $\beps$ and the syndrome $\bH_v\be_v^\intercal$, where now $\be_v$ denotes the restriction of the error vector $\be_X$ to the local view $Q(v)$ and $\bH_v$ denotes the restriction of the parity-check matrix $\bH_{Z}$ to the set of constraints imposed by the vertex $v.$

Denote by $\bu_X\in\F_2^N$ the vector that is identical to $\be_X$ on the coordinates $i\in[N]$ that we were able to recover and has zeroes everywhere else.  Denote by $\bu_Z\in\F_2^N$ the vector that is identical to $\be_Z$ on the coordinates $i\in[N]$ that we were able to recover and has zeroes everywhere else. As the final step of our decoding algorithm, apply the adversarial algorithm of Theorem \ref{thm:decoding} (\cite{LZ2023}) to the $X$-syndrome 
$\bH_X\be_Z^\intercal+\bH_X\bu_Z^\intercal$ and the $Z$-syndrome $\bH_Z\be_X^\intercal+\bH_Z\bu_X^\intercal$.

\an{\textbf{ChatGPT comment.}
Replace the proof opening and algorithm description through the preceding residual-syndrome sentence by:
\begin{quote}
Let $\beps\in\{0,1\}^N$ be the erasure pattern, and let $\be_X,\be_Z$ be arbitrary vectors supported in $\supp(\beps)$. We are given $s_X=\bH_X\be_Z^\intercal$ and $s_Z=\bH_Z\be_X^\intercal$, and seek representatives of $\be_Z+C_Z^\perp$ and $\be_X+C_X^\perp$.
For each $v\in V_{00}$, use the local $X$-check syndrome and the restricted erasure pattern to solve for $\be_Z|_{Q(v)}$. Accept the solution only if the matrix restricted to erased columns has full column rank; otherwise leave this block unresolved. Let $\bu_Z$ equal the recovered entries on successful blocks and zero on all other coordinates. Similarly, use the local $Z$-check syndromes at $v\in V_{01}$ to recover $\be_X$ on successful blocks and form $\bu_X$.
Apply the adversarial decoder to the residual syndromes $s_X+\bH_X\bu_Z^\intercal$ and $s_Z+\bH_Z\bu_X^\intercal$. If it returns residual corrections $\br_Z,\br_X$, output $\bu_Z+\br_Z$ and $\bu_X+\br_X$, respectively. These represent the required stabilizer cosets whenever the residual weights are within its guaranteed radius.
\end{quote}
The replacement makes the full-column-rank acceptance test explicit and adds the recovered vectors back to the residual corrections. Keep $N_X$ for failed $X$-check blocks at $V_{00}$ and $N_Z$ for failed $Z$-check blocks at $V_{01}$, but define them as failures of the full-column-rank test; this makes them depend only on the erasure pattern. In the Gaussian-elimination footnote the unerased error entries are zero, so the right-hand side is simply the local syndrome.}

We first bound the probability that our algorithm fails. Define
\begin{align*}
    N_X:=\Big|\big\{v\in V_{00}:\be_v \textnormal{ cannot be recovered from $\beps$ and }\bH_v\be_v^\intercal \big\} \Big|,
\end{align*}
\begin{align*}
    N_Z:=\Big|\big\{\tilde{v}\in V_{01}:\be_{\tilde{v}} \textnormal{ cannot be recovered from $\beps$ and }\bH_{\tilde{v}}\be_{\tilde{v}}^\intercal \big\} \Big|.
\end{align*}
Here the subscripts of $N_X$ and $N_Z$ denote the check type: $N_X$ counts failed local recoveries of $\be_Z$ at $V_{00}$, and $N_Z$ counts failed local recoveries of $\be_X$ at $V_{01}$.
By Corollary \ref{cor:decoding}, as long as $\Delta$ is large enough in terms of $\rho,\varepsilon$, there exists a constant $f(\rho,\varepsilon)$ such that whenever the first part of our algorithm manages to recover all but $\frac{f(\rho,\varepsilon)N}{\Delta}$ erroneous entries of $\be_X$ and $\be_Z$, 
%. If $N_X<f(\rho,\varepsilon)N/(2\Delta^3)$ and $N_Z<f(\rho,\varepsilon)N/(2\Delta^3)$, the sum of the two residual component weights is less than $f(\rho,\varepsilon)N/\Delta$ and
the adversarial decoding of \cite{LZ2023} is guaranteed to succeed.
Thus, since every vector $\be_v$ has length $\Delta^2$, we have that for all $\Delta$ large enough in terms of $\rho$ and $\varepsilon$,
%\begin{align*}
   % \Pr_{\substack{E\sim p\\e_X,e_Z\sim \mu(E)}}\left[ \textnormal{decode}(C,E,e_X,e_z) =0\right]&\leq \Pr_{\substack{E\sim p\\e_x\sim \mu(E)}}\left[N_{X}\geq \frac{f(\rho,\varepsilon)n}{2\Delta^3}\right]+\Pr_{\substack{E\sim p\\e_Z\sim \mu(E)}}\left[N_{Z}\geq \frac{f(\rho,\varepsilon)n}{2\Delta^3}\right]+2^{-\Omega(n)}
%\end{align*}
\begin{align}\label{PrErr}
    \Pr_{}\Big[ \textnormal{decode}(C_{\rho,\varepsilon,\Delta,N},\beps) =0\Big]&\leq \Pr_{}\left[N_{X}\geq \frac{f(\rho,\varepsilon)N}{2\Delta^3}\right]+\Pr_{}\left[N_{Z}\geq \frac{f(\rho,\varepsilon)N}{2\Delta^3}\right]. %+2^{-bN}
\end{align}
%for some constant $b=b(\rho,\varepsilon).$

%\an{\textbf{ChatGPT comment.}
%Theorem~\ref{thm:decoding} corrects every error within its radius; no additional random failure term is stated. Replace the sentence saying it succeeds ``with high probability'' by:
%\begin{quote}
%If $N_X<f(\rho,\varepsilon)N/(2\Delta^3)$ and $N_Z<f(\rho,\varepsilon)N/(2\Delta^3)$, the sum of the two residual component weights is less than $f(\rho,\varepsilon)N/\Delta$, so the adversarial decoder is guaranteed to succeed.
%\end{quote}
%Replace \eqref{PrErr} by
%\[
% \begin{aligned}
% &\Pr_{\beps\sim\rho_N}[D_N\text{ fails for some supported Pauli error}]\\
% &\quad\leq\Pr[N_X\geq f(\rho,\varepsilon)N/(2\Delta^3)]\\
% &\qquad+\Pr[N_Z\geq f(\rho,\varepsilon)N/(2\Delta^3)].
% \end{aligned}
%\]
%Delete the following sentence introducing $b$. This also replaces the undefined four-argument expression $\textnormal{decode}(C,\beps,\be_X,\be_z)$ and the inconsistent lowercase subscript $z$. After fixing the local codes, disjoint local blocks give the independence required for both Chernoff bounds. Independence between $N_X$ and $N_Z$ is unnecessary; a union bound suffices.}
Now for any vertex $v\in V_{00}$, we succeed in recovering $\be_{v}$ from $\beps$ and $\bH_{v}\be_{v}^\intercal$ if and only if the restriction of $\bH_v$ to the columns indexed by $\beps$ has full rank (see footnote \ref{footnotealgo}), which by Lemma \ref{equivalentformulationscoveredsubcode} happens if and only if the classical code $(C_A\otimes C_B)^\perp$ is capable of decoding the error string $\beps$. But recall from equation \eqref{firstgood} that
\begin{align*}%\label{firstgood}
\Pr_{\substack{x\sim \rho+\frac{\varepsilon}{15}}}\Big[\mathsf{decode}\left((C_A\otimes C_B)^\perp,x\right)=0\Big]&\leq \Delta^{-\frac{1}{10}}.%,
\end{align*}
%$$\textnormal{rate}(C_A\otimes C_B)\geq\rho +\frac{\varepsilon}{12}.$$ %and that the code used for the $V_{01}$ checks has rate
%$$\textnormal{rate}(C_A^\perp\otimes D_B^\perp)\geq\frac{\varepsilon}{12}.$$
Thus by Theorem \ref{sharptransition}, there exists a constant $\alpha>0$ for which
\begin{align*}
\Pr_{\substack{\bx\sim \rho+\frac{{\varepsilon}}{20}}}\Big[\mathsf{decode}\left((C_A\otimes C_B)^\perp,\bx\right)=0\Big]&\leq e^{-\alpha \varepsilon^2\Delta}\leq \frac{f(\rho,\varepsilon)}{4\Delta},
\end{align*}
for $f(\rho,\varepsilon)$ the constant in Corollary \ref{cor:decoding} and $\Delta$ large enough in terms of $\rho,\varepsilon$. %In the exact same way, we have

\an{\textbf{ChatGPT comment.}
Theorem~\ref{thmrandomduals} gives an ensemble average. It cannot imply the displayed $1/2$ bound for every local-code realization, nor can one apply the fixed-code sharp-transition theorem directly to that average. Moreover all global vertices reuse the same tuple of local codes, so their failures are independent only after fixing that tuple. Replace the argument from ``For $\Delta$ large enough'' through the preceding display by:
\begin{quote}
Put $p_+=\rho+\eta/24$. For each of the two active local dual tensor codes, Theorem~\ref{thmrandomduals} gives ensemble-average failure probability $o(1)$ at $p_+$ as $\Delta\to\infty$, using the margins in \eqref{goodrateX}--\eqref{goodrateZ}. Markov's inequality implies that, with probability $1-o(1)$ over the four component codes, each of these two fixed-code failure probabilities is at most $1/2$. Simultaneously, all required distance and product-expansion properties hold with probability $1-o(1)$. A union bound therefore allows us to choose one tuple satisfying every requirement, and we fix that tuple independently of the global length $N$.
For either active local dual tensor code $L$, product expansion gives $d_{\min}(L)\geq\kappa\Delta$: every nonzero word requires at least one nonzero constituent row or column in its decomposition. Its median erasure threshold is at least $p_+$. Theorem~\ref{sharptransition} therefore yields
\[
 \Pr_{\bx\sim\rho_{\Delta^2}}[\mathsf{decode}(L,\bx)=0]
 \leq\exp\bigl(-a_0\kappa(\eta/24)^2\Delta\bigr)
 \leq\frac{f(\rho,\varepsilon)}{4\Delta}
\]
for sufficiently large $\Delta$, where $a_0>0$ is the universal constant in that theorem. All local choices are made before the Cayley complexes grow.
\end{quote}
The effective coefficient in an expression $e^{-\alpha\varepsilon^2\Delta}$ is not universal: it includes the local distance or expansion constant and therefore depends on the chosen rates and gap. The replacement makes that dependence explicit and removes the source of a nonvanishing ensemble-error floor in the claimed global exponent.}
%\begin{align*}
%\Pr_{z\sim \rho}\Big[\mathsf{decode}\left((C_A^\perp\otimes D_B^\perp)^\perp,z\right)=0\Big]&\leq \frac{f(\rho,\varepsilon)}{4\Delta}.
%\end{align*}
Thus by the Chernoff bound, since on average we expect $\leq\frac{N}{\Delta^2}\cdot \frac{f(\rho,\varepsilon)}{4\Delta}= \frac{f(\rho,\varepsilon)N}{4\Delta^3}$ vertices of $V_{00}$ to have a decoding failure and since the local views of any two distinct vertices $v,v'\in V_{00}$ are disjoint, we have
\begin{align}\label{bdX}
    \Pr\left[N_{X}\geq \frac{f(\rho,\varepsilon)N}{2\Delta^3}\right]\leq e^{-\frac{f(\rho,\varepsilon)N
    }{12\Delta^3}} .
\end{align}
The exact same argument applied to the $Z$-checks of $V_{01}$ and the code $C_A^\perp\otimes D_B^\perp$ %(whose rate is also no smaller than $\rho+\frac{\varepsilon}{12}$, as shown in equation \eqref{goodrateZ}) 
gives
\begin{align}\label{bdZ}
    \Pr\left[N_{Z}\geq \frac{f(\rho,\varepsilon)N}{2\Delta^3}\right]\leq e^{-\frac{f(\rho,\varepsilon)N}{12\Delta^3}}.
\end{align}
Fixing some $\Delta$ and combining \eqref{bdX} and\eqref{bdZ} with equation \eqref{PrErr}, we get that there exists a constant $a=a(\rho,\varepsilon)$ for which
\begin{align*}
    \Pr_{\substack{\beps\sim\rho}}\Big[ \textnormal{decode}(C_{\rho,\varepsilon,\Delta,N},\beps) =0\Big]&\leq 2^{-aN}
\end{align*}
for all $N$ large enough, as desired. All that remains to show is then that the runtime of our algorithm is linear in $N$. For any $v\in V_{00}\cup V_{01}$, the vector $\be_v$ has length $\Delta^2$ and the matrix $\bH_v$ has fewer than $\Delta^2$ rows. Since our chosen $\Delta$ is dependent only on $\rho$ and $\varepsilon$, the Gaussian elimination process described in footnote \ref{footnotealgo} runs in time independent on $N$, and decoding $\be_v$ for all the vertices $v\in V_{00}\cup V_{01}$ thus takes time $\leq \beta N$ for some constant $\beta=\beta(\varepsilon,\rho).$ Then since each row of $\bH_X,\bH_Z$ has weight at most $\Delta^2=O_{\varepsilon,\rho}(1)$, constructing $\bH_X\bu_Z^\intercal$ and $\bH_Z\bu_X^\intercal$ (and summing them with $\bH_X\be_Z^\intercal$ and $\bH_Z\be_X^\intercal$ respectively) takes time linear in $N$ as well. Finally, by Corollary \ref{cor:decoding}, the adversarial algorithm of \cite{LZ2023} takes time $O(N)$. Thus overall, our algorithm runs in linear time.

%for all $N$ large enough, as desired. All that remains to show is then that the runtime of our algorithm is linear in $N$. For any $v\in V_{00}\cup V_{01}$, the vector $\be_v$ has length $\Delta^2$ and the matrix $\bH_v$ has fewer than $\Delta^2$ rows. The Gaussian elimination process described in footnote \ref{footnotealgo} thus runs in time $O(\Delta^6)$, and decoding $\be_v$ for all the vertices $v\in V_{00}\cup V_{11}$ takes time $O(\Delta^4 N).$ For our fixed $\Delta$ (i.e. dependent only on $\rho$ and $\varepsilon$ but not on $N$) Then since each row of $\bH_X,\bH_Z$ has weight $\leq\Delta^2$, constructing $\bH_X\bu_X^\intercal$ and $\bH_Z\bu_Z^\intercal$ (and summing them with $\bH_X\be_X^\intercal$ and $\bH_Z\be_Z^\intercal$ respectively) takes time $O(\Delta^2 n).$ Finally, by Corollary \ref{cor:decoding}, the adversarial algorithm of \cite{LZ2023} takes time $O(n)$ DEPENDENCE ON $\Delta??$. Thus overall, our algorithm takes time $O(\Delta^4 n).$
\end{proof}

\section*{Acknowledgements}
The work of Henry D. Pfister was supported in part by the NSF under Grant 2120757. The work of Oscar
Sprumont and of Gilles Z{\'{e}}mor was supported by Plan France 2030 through the Project
NISQ2LSQ, ANR-22-PETQ-0006. A conversation with the free version of ChatGPT helped speed up the proof of Lemma \ref{correlationshurt}.
{\color{\markcolor} While the manuscript was drafted without artificial intelligence, ChatGPT was used for editing assistance and to help find typos and errors.}
%Artificial intelligence was not used in any part of the writing process.
The authors take full responsibility for the contents of this paper.

\bibliographystyle{alpha}
\bibliography{biblio}

\end{document}